\pdfoutput=1
\documentclass[10pt,journal,compsoc]{IEEEtran}

\ifCLASSOPTIONcompsoc
\else
  \usepackage{cite}
\fi

\ifCLASSINFOpdf
  \usepackage[pdftex]{graphicx}
  \graphicspath{{figs}}
  \DeclareGraphicsExtensions{.pdf,.jpeg,.png}
\else
\fi

\usepackage{amsmath}

\usepackage{array}

\ifCLASSOPTIONcompsoc
  \usepackage[caption=false,font=footnotesize,labelfont=sf,textfont=sf]{subfig}
\else
  \usepackage[caption=false,font=footnotesize]{subfig}
\fi

\usepackage{fixltx2e}

\usepackage{dblfloatfix}

\ifCLASSOPTIONcaptionsoff
  \usepackage[nomarkers]{endfloat}
 \let\MYoriglatexcaption\caption
 \renewcommand{\caption}[2][\relax]{\MYoriglatexcaption[#2]{#2}}
\fi

\usepackage{url}

\usepackage{color}
\usepackage[formats]{listings}

\usepackage{stmaryrd}

\ifCLASSOPTIONcompsoc
      \usepackage{amssymb,amstext}
\else
      \usepackage{graphicx}
      \usepackage{amsmath,amssymb,amstext} %
\fi

\usepackage{xcolor}
\usepackage[
    verbose,
    colorlinks,
    linkcolor={blue!50!black},
    citecolor={blue!50!black},
    urlcolor={blue!80!black}
]{hyperref}

\makeatletter

\usepackage{xfp}
\usepackage{fmtcount}

\usepackage{afterpage}

\newcommand{\numexpert}{74}

\newcommand{\numqueriestotal}{364}
\newcommand{\numqueriesportuscantdosecondorder}{14} %
\newcommand{\numqueriesportuscantdounivfield}{7}
\newcommand{\numqueriesportuscantdobadjoin}{5}
\newcommand{\numquerieskodkodcantdo}{1}

\newcommand{\numqueriesportuscantdonondefinitesort}{\fpeval{\numqueriesportuscantdounivfield + \numqueriesportuscantdobadjoin}}

\newcommand{\numqueriesportuscantdo}{\fpeval{\numqueriesportuscantdosecondorder + \numqueriesportuscantdonondefinitesort}}

\newcommand{\totalcorrectnessqueries}{\fpeval{\numqueriestotal - \numqueriesportuscantdo - \numquerieskodkodcantdo}}
\newcommand{\numquerieswithsecondorder}{3} %
\newcommand{\numqueriesportustimeout}{30} %

\newcommand{\numcorrectnesssatunfinished}{6} %
\newcommand{\numcorrectnessunsatunfinished}{24} %

\newcommand{\numcorrectnesssat}{183} %
\newcommand{\numcorrectnessunsat}{124} %
\newcommand{\totalcorrectnessqueriesfinished}{\fpeval{\numcorrectnesssat + \numcorrectnessunsat}}
\newcommand{\totalcorrectnessqueriessat}{\fpeval{\numcorrectnesssat + \numcorrectnesssatunfinished}}
\newcommand{\totalcorrectnessqueriesunsat}{\fpeval{\numcorrectnessunsat + \numcorrectnessunsatunfinished}}

\newcommand{\numsupport}{63}

\newcommand{\numsat}{42}
\newcommand{\numunsat}{\fpeval{\numsupport - \numsat}} %

\newcommand{\toolurl}{\texttt{\href{https://github.com/WatForm/org.alloytools.alloy/tree/portus}{https://github.com/WatForm/org.alloytools.alloy}} (branch \texttt{portus})}

\newcommand{\portus}{\textsc{Portus}}
\newcommand{\kodkod}{\textsc{Kodkod}}
\newcommand{\fortress}{\textsc{Fortress}}
\newcommand{\alloytosmt}{\textsc{alloy2smt}}
\newcommand{\fortresslang}{F-MSFOL}
\newcommand{\zthree}{\textsc{Z3}}
\newcommand{\cvcfive}{\textsc{CVC5}}
\newcommand{\cvcfour}{\textsc{CVC4}}
\newcommand{\alloyanalyzer}{\textsc{Alloy Analyzer}}
\newcommand{\satj}{\textsc{Sat4j}}
\newcommand{\minisat}{\textsc{MiniSat}}
\newcommand{\alloype}{\textsc{AlloyPE}}
\newcommand{\prioni}{\textsc{Prioni}}
\newcommand{\qalloy}{\textsc{QAlloy}}
\newcommand{\crs}{\textsc{CRS}}
\newcommand{\kelloy}{\textsc{Kelloy}}
\newcommand{\key}{\textsc{KeY}}
\newcommand{\dynamite}{\textsc{Dynamite}}
\newcommand{\pvs}{\textsc{PVS}}
\newcommand{\allealle}{\textsc{AlleAlle}}
\newcommand{\vampire}{\textsc{Vampire}}
\newcommand{\nusmv}{\textsc{NuSMV}}
\newcommand{\nuxmv}{\textsc{nuXmv}}
\newcommand{\prob}{\textsc{ProB}}
\newcommand{\alloytojml}{\textsc{Alloy2JML}}

\newcommand{\myetal}{et al.}
\newcommand{\myeg}{e.g.,}
\newcommand{\myie}{i.e.,}
\newcommand{\myetc}{etc.}
\newcommand{\dfn}[1]{\textbf{#1}}

\newcommand{\better}[1]{}

\newcommand{\drop}[1]{}

\newcommand{\qand}{\wedge}
\newcommand{\qor}{\vee}
\newcommand{\qimplies}{\Rightarrow}
\newcommand{\qnot}{\neg}
\newcommand{\qiff}{\Leftrightarrow}

\newcommand{\qforall}{\forall}
\newcommand{\qexists}{\exists}
\newcommand{\st}{\bullet}
\newcommand{\qunion}{\cup}

\newcommand{\qite}[3]{#1\mathrel{\mathord{?}}#2\mathrel{\mathord{:}}#3}

\newcommand{\qceil}[1]{{\lceil}#1{\rceil}}

\newcommand{\trans}[1]{\llbracket #1 \rrbracket}

\newcommand{\arity}[1]{arity(#1)}

\lstdefinelanguage{MyAlloy}{
morekeywords={sig, fact, assert, one, extends, some, all, iff, set, or, and, not, in, abstract, fun, let, module, open, check, exactly, for, pred, run, lone, one, no, implies, sum, disj, univ, none, iden},
sensitive=true,
morecomment=[l]{//},
morecomment=[s]{/*}{*/}
}
\newcommand{\alloymath}[1]{%
    \hbox{\mbox{\lstinline[language=MyAlloy,basicstyle=\ttfamily,columns=fixed]{#1}}}}

\newcommand{\alloy}[1]{%
    \lstinline[language=MyAlloy,basicstyle=\ttfamily,columns=fixed]{#1}}

\newcommand{\alloytext}[1]{\text{\alloy{#1}}}

\newcommand{\fortressuniv}{\mathit{univ}}
\newcommand{\fortressint}{\mathit{Int}}
\newcommand{\fortressbool}{\mathit{Bool}}

\newcommand{\fname}[1]{\mathit{#1}}
\newcommand{\aux}{\fname{aux}}
\newcommand{\sortfn}{\mathsf{sortpolicy}} %
\newcommand{\resolvant}{\mathsf{resolvant}}
\newcommand{\casttoscalar}{\mathsf{castToScalar}}

\newcommand{\resques}[1]{\textbf{Research Question: \textit{#1}}}

\begin{document}
\title{\portus: Linking Alloy with SMT-based Finite Model Finding}

\author{Ryan~Dancy, Nancy~A.~Day,~\IEEEmembership{Member,~IEEE Computer Society}, Owen~Zila, Khadija~Tariq, Joseph~Poremba%
  \IEEEcompsocitemizethanks{%
    \IEEEcompsocthanksitem Ryan Dancy (rdancy@uwaterloo.ca), Nancy A.\ Day (nday@uwaterloo.ca), Owen Zila (owen.zila@uwaterloo.ca), and Khadija~Tariq (k2tariq@uwaterloo.ca) are with the David R. Cheriton School of Computer Science, University of Waterloo.%
    \IEEEcompsocthanksitem Joseph Poremba (jporemba@cs.ubc.ca) is with the Department of Computer Science, University of British Columbia.%
  }%
\thanks{© 2025 IEEE. Personal use of this material is permitted. Permission from IEEE must be obtained for all other uses, in any current or future media, including reprinting/republishing this material for advertising or promotional purposes, creating new collective works, for resale or redistribution to servers or lists, or reuse of any copyrighted component of this work in other works.}}

\markboth{}%
{}

\IEEEtitleabstractindextext{%
Alloy is a well-known, formal, declarative language for modelling systems early in the software development process. Currently, it uses the \kodkod\ library as a back-end for finite model finding. \kodkod\ translates the model to a SAT problem; however, this method can often handle only problems of fairly low-size sets and is inherently finite.  We present \portus, a method for translating Alloy into an equivalent many-sorted first-order logic problem (MSFOL).  Once in MSFOL, the problem can be evaluated by an SMT-based finite model finding method implemented in the \fortress\ library, creating an alternative back-end for the \alloyanalyzer.  \fortress\ converts the MSFOL finite model finding problem into the logic of uninterpreted functions with equality (EUF), a decidable fragment of first-order logic that is well-supported in many SMT solvers. We compare the performance of \portus\ with 
\kodkod\ on a corpus of \numsupport\ Alloy models written by experts.  
Our method is fully integrated into the \alloyanalyzer.

}

\maketitle

\IEEEdisplaynontitleabstractindextext

\IEEEpeerreviewmaketitle

\section{Introduction}

Alloy~\cite{Jackson2012} is a declarative modelling language suitable for describing systems early in the development process.
The model is described using %
relations and constraints over these relations.  
Critical to the value of early modelling is the quick feedback from finite model finding to help find bugs prior to a huge investment of time in specification.  
Finite model finding (FMF) means searching for satisfying instances of a finite size of a formal model, thus making it a decidable problem.  
Finite model finding of Alloy models has been used to formally analyze many interesting case studies in the literature (\myeg\ ~\cite{Zave2012a, Cunha2020, deKinderen2023, Zeroual2023, Benavides2022, Jahanian2020, Cunha2024}).

 The \alloyanalyzer\ currently translates an FMF problem into 
a propositional logic satisfiability problem via the \kodkod~\cite{Torlak2007} library, which links to a variety of SAT solvers.
While this method gives quick feedback for small size (scope) sets, it reaches capacity limitations as scopes increase and only works for completely finite problems.  Vakili and Day~\cite{Vakili2016b} show that the FMF problem for many-sorted, first-order logic (MSFOL) can be expressed in the logic of uninterpreted functions with equality (EUF), which is a decidable subset of MSFOL~\cite{Ackermann1962} that is well-supported in SMT (satisfiability modulo theories) solvers~\cite{Barrett2018}.  SMT solvers may be able to solve a finite model finding problem more quickly than a SAT solver because the structure of functions 
can be exploited during solving.  Vakili and Day implement their approach in a library called \fortress, which links with various SMT solvers.
SMT solvers can solve both finite and unbounded problems, which gives them more flexibility compared to \kodkod.

We present a method, called \portus, for translating Alloy models into MSFOL. Our translation allows us to connect the \alloyanalyzer\ with the \fortress\ library.  
Compared to related efforts to link Alloy with SMT solvers 
(\myeg~\cite{ElGhazi2010,ElGhazi2011a, ElGhazi2014,Meng2017,Mohamed2019}), our
goal in this work is to use the SMT solver for finite model finding (via a mapping to EUF) rather than to solve for unbounded scopes.  Our translation includes novel methods for handling scalars and functions, non-exact scopes, set hierarchy (through sort policies), and set cardinality. Our method covers 
the many subtleties of the Alloy signature hierarchy to create an almost-complete translation of everything in the Alloy language.
In future work, we plan to exploit the flexibility of the SMT solver for analyzing a hybrid of bounded and unbounded scope problems.

We begin by presenting a general translation that maps Alloy signatures to MSFOL predicates,\footnote{Everything in MSFOL is a function.  A predicate is function with a Boolean result.} and maps set/relation operations to their counterparts over predicates in MSFOL.   Our translation
uses definitions and tries to avoid quantifiers as much as possible to gain performance in 
\fortress{}.
Next,
we discuss optimizations that exploit the features of MSFOL (and SMT solvers) such as multiple sorts, scalars, and functions. Then, we evaluate 
the completeness and correctness of our implementation 
on a Benchmark Set 
of models written by experts, scraped from the web by Eid and Day~\cite{Eid2023}. 
Finally, we compare the performance of our optimizations and the performance of \portus\ versus \kodkod\ on queries from the Benchmark Set.  We find that \portus\ is competitive with the 
performance of \kodkod\ and for hard signatures of some models, it scales 
better than \kodkod.
No solver will be best for all problems~\cite{Wang2019a}, so \portus\ adds an additional solver that performs better on some Alloy models. 
Our solver\footnote{\toolurl} is directly integrated into the \alloyanalyzer\ 6 including the presentation of instances discovered by the solver within the Alloy evaluator.

\section{Background}

\textbf{Alloy}~\cite{Jackson2012} is a flexible modelling language based on relational logic.
It represents models using sets (signatures) and relations, and constraints on these sets/relations are expressed in a  relational calculus. %
Alloy's signatures can be arranged in a hierarchy. %
The Alloy language does not distinguish between sets and scalars, so only a small set of operators is needed---every 
value of a signature is considered as a set of atoms.  The value of a relation is a set of tuples of atoms.
All sets are flat---there are no sets of sets, which allows us to map it to MSFOL, where a set can be described via a membership predicate. 
An Alloy model consists of signature declarations (which introduce sets and relations and may include formulas), formulas, and commands.
Alloy includes the second-order operators of transitive closure and set cardinality.  
More details about the Alloy modelling language are provided in later sections as needed.

The \alloyanalyzer\ can be used either to find a sample instance of a model using the \alloy{run} command or to check if the model violates a given property and produce a counterexample using the \alloy{check} command.
We refer to the combination of an Alloy model and a single command as a \textbf{query}.
This analysis is fully automated but it only checks queries for solutions consisting of signatures of size within a finite bound (called a \textbf{scope}). 
\kodkod~\cite{Torlak2007} is a library for finite model finding for a more basic relational logic than Alloy. The \alloyanalyzer\ converts an Alloy model to an untyped \kodkod\ problem,
which \kodkod\ translates to propositional logic and uses off-the-shelf SAT solvers for evaluation. 
Second-order operators such as transitive closure and set cardinality are expressed by fully expanding their meaning for the finite scope. Particular effort has been made in \kodkod\ to do symmetry reductions.

\textbf{\fortress}~\cite{Vakili2016b} is a library %
that takes as input a many-sorted first-order logic (MSFOL) finite model finding problem, called a \textbf{theory}, with finite scopes for the sorts. 
In MSFOL, the sets of values for the sorts must be disjoint.
\fortress{} converts the theory to the logic of equality and uninterpreted functions (EUF), and checks its satisfiability by invoking an SMT solver.  We use standard logical notation for MSFOL including $\top$ (true) and $\bot$ (false).
\fortress\
adds \textit{range formulas} for all constant and function symbols to constrain their output values to the finite set of possible sort elements, which forces the solver to consider only finite interpretations of a certain size of the problem.  \fortress\ introduces distinct constants (called \textbf{domain elements}) to represent the sort elements, normalizes and skolemizes the formulas, grounds the formulas by instantiating the universal quantifiers with the domain elements, and introduces range formulas.  %
If the FMF problem is satisfiable, \fortress\ can retrieve an interpretation for the constants and functions.
\fortress\ has an extensive set of symmetry breaking schemes to reduce the problem state space~\cite{Poremba2023}. 
For integers included in models, \fortress\ uses a method called overflow-preventing finite integers (OPFI)~\cite{Zila2023} to reduce overflow errors. OPFI uses unbounded integers (which is possible in SMT solvers) for intermediate values of a term, while continuing to limit the range of exploration to finite integers for decidability. 
We use the name \textbf{\fortresslang} for the input language of \fortress\ because it extends MSFOL to include (reflexive) transitive closure as a predicate that takes a binary predicate as an argument.
Additionally, \fortresslang\ includes second-order quantifiers and \fortress\ is able to skolemize existential quantification.
\fortress{} supports trailing arguments for closures. 
$\fname{Closure}(p, x, y, a_1, \dots, a_n)$ is true if $(x,y)$ is in the closure of predicate $p$, where $x,y$ are the arguments to $p$ which are closed over and $a_1,\dots,a_n$ are extra values that are passed to every application of $p$.
The \fortress\ library transforms the transitive closure operators 
into MSFOL through the iterative squaring approach of Burch \myetal~\cite{Burch1992}, where 
for a domain of size $k$, $\qceil{\log{k}}$ auxiliary definitions are used to define a function representing the transitive closure of a predicate.

Edwards, Jackson, and Torlak~\cite{Edwards2004a} present a bounding type system used for a form of typechecking in Alloy. 
Their type system assigns a type to each Alloy expression \alloy{e}. A type is a set of tuples of signatures such that every tuple in an interpretation of \alloy{e} is guaranteed to lie in one of the tuples of signatures in its type.
In Section~\ref{sec:sort-policy}, we present a sort resolvant system to match Alloy signatures to \fortresslang{} sorts for our translation that builds on 
the type system of ~\cite{Edwards2004a}.
However, \cite{Edwards2004a} assigns a distinct atomic type to each signature; in contrast, our assignment of sorts is coarser and is based on a `sort policy', as explained in Section~\ref{sec:sort-policy}.
In particular, we do not use the `remainder types' of \cite{Edwards2004a}.
When the partition sort policy of Section~\ref{sec:partition-sort-policy} is in use, our atomic types generally correspond to top-level signatures.

\section{General Translation}
\label{sec:basic}

This section describes our general, unoptimized method to translate an Alloy model to \fortresslang.
As input, \portus{} receives from the \alloyanalyzer\ a list of the signatures, fields, facts (including the formula of the selected \alloy{run} command or the negated formula of a \alloy{check} command), and function and predicate definitions present in the model, along with the scope of each signature\footnote{The Alloy language allows a model to be split into several modules.
The Analyzer loads and performs relevant substitutions on the imported modules prior to our translation step, so our process does not need any special support for modules.}.
From this, \portus{} produces as output a \fortress{} theory including sorts, constant and function declarations and definitions, and 
axioms.

We define a recursive translation operator, $\trans{\cdot}$, over the abstract syntax tree (AST) for Alloy terms, which maps an Alloy term to an equivalent \fortresslang\ formula.
The translation process may have side effects: for example, translating an Alloy formula may cause auxiliary functions to be added to the \fortress{} theory.
The general translation proceeds in three steps: 1) translate the signatures (including the signature hierarchy) into declarations and axioms in \fortresslang; 2) translate fields into \fortresslang\ functions limited by axioms; and 3) translate the facts to \fortresslang\ formulas via a top-down traversal.
Frequently, we define the translation of one Alloy expression in terms of the translation of another, simpler Alloy expression.  This method makes our core translations a smaller set and allows us to apply optimizations (Section~\ref{sec:opt})
more generally.

A set of Alloy atoms cannot be directly represented in \fortresslang{}, which can only represent scalars (individual values rather than sets of values).
Therefore, we must translate $(x_1,\dots,x_n) \in \alloytext{e}$ rather than the Alloy expression \alloy{e} directly.
In the following, we use the syntax $(x_1,\dots,x_n) \in \alloytext{e}$ as an argument to the translation function, even though it is not part of the Alloy syntax, to mean that the tuple of Alloy atoms corresponding to the \fortresslang\ atoms $(x_1,\dots,x_n)$ is a member of the set generated by the Alloy expression \alloy{e}, and we show its translation.
The \fortresslang{} sorts to which $x_1,\dots,x_n$ belong are passed along with the tuple.

During translation, it is necessary to know the \fortresslang\ sorts corresponding to an Alloy expression, such as for the purpose of quantifying over the expression.
We begin this section by discussing our sort resolvant function and sort policies, which determine the sort of an \fortresslang\ formula.

Our general translation builds on translation parts from related work (particularly~\cite{ElGhazi2010, ElGhazi2011a}) such as membership predicates, axioms for subsignatures, translations of logical operations and many set operations so these are covered very briefly here for completeness.  Section~\ref{sec:related} highlights the novelty in our overall approach (general translation plus optimizations) and provides a full comparison between our work and related work.

\subsection{Sort Resolvants and Sort Policy}
\label{sec:sort-policy}

\begin{figure*}[h]
\begin{align*}
  \resolvant(\textrm{\alloy{s}}) & := \{(\sortfn(\textrm{\alloy{s}}))\} \tag*{for all sigs \alloy{s}}  \\
  \resolvant(\textrm{\alloy{f}}) & := \{(\sortfn(\textrm{\alloy{s}}))\} \times \resolvant(\textrm{\alloy{e}}) \tag*{for fields \alloy{f: e} within sig \alloy{s}}  \\
  \resolvant(\textrm{\alloy{x}}) & := \{(\sortfn(\textrm{\alloy{e}}))\} \tag*{for quantified variables \alloy{x: e}} \\
  \resolvant(\textrm{\alloy{e1 + e2}}) & := \resolvant(\textrm{\alloy{e1}}) \cup \resolvant(\textrm{\alloy{e2}}) \\
  \resolvant(\textrm{\alloy{e1 \& e2}}) & := \resolvant(\textrm{\alloy{e1}}) \cap \resolvant(\textrm{\alloy{e2}})  \\
  \resolvant(\textrm{\alloy{e1 - e2}}) & := \resolvant(\textrm{\alloy{e1}}) \\
  \resolvant(\textrm{\alloy{e1 ++ e2}}) & := \resolvant(\textrm{\alloy{e1}}) \cup \resolvant(\textrm{\alloy{e2}})  \\
  \resolvant(\textrm{\alloy{e1 <: e2}}) & := \{(S_1, \dots, S_m) \in \resolvant(\textrm{\alloy{e2}}) : (S_1) \in \resolvant(\textrm{\alloy{e1}})\} \\
  & \qquad \text{where $m = \arity{\textrm{\alloy{e2}}}$}\\
  \resolvant(\textrm{\alloy{e1 :> e2}}) & := \{(S_1, \dots, S_n) \in \resolvant(\textrm{\alloy{e1}}) : (S_n) \in \resolvant(\textrm{\alloy{e2}})\} \\
  & \qquad \text{where $n = \arity{\textrm{\alloy{e1}}}$}\\
  \resolvant(\textrm{\alloy{e1 . e2}}) & := \{(S_1, \dots, S_{n-1}, S'_2, \dots, S'_m) : (S_1, \dots, S_n) \in \resolvant(\textrm{\alloy{e1}}) \\
  & \qquad ~~\text{and } (S'_1, \dots, S'_m) \in \resolvant(\textrm{\alloy{e2}}) \text{ and } S_n = S'_1 \} \\
  & \qquad \text{where $n = \arity{\textrm{\alloy{e1}}}$ and $m = \arity{\textrm{\alloy{e2}}}$} \\
  \resolvant(\textrm{\alloy{e1 M->N e2}}) & := \resolvant(\textrm{\alloy{e1}}) \times \resolvant(\textrm{\alloy{e2}}) \tag*{for all multiplicities \alloy{M}, \alloy{N}}  \\
  \resolvant({\textrm{\texttt{\textasciitilde}\alloy{e}}}) & := \{(S_2, S_1) : (S_1, S_2) \in \resolvant(\textrm{\alloy{e}})\}  \\
  \resolvant(\textrm{\alloy{cond => e1 else e2}}) & := \resolvant(\textrm{\alloy{e1}}) \cup \resolvant(\textrm{\alloy{e2}}) \qquad \\
  \resolvant(\textrm{\alloy{\{x1: e1, ..., xn: en | ...\}}}) & := \resolvant(\textrm{\alloy{e1}}) \times \dots \times \resolvant(\textrm{\alloy{en}})  \\
  \resolvant({\textrm{\texttt{\textasciicircum}\alloy{e}}}) & := \resolvant(\textrm{\alloy{e}})  \\
  \resolvant(\textrm{\alloy{*e}}) & := \resolvant(\textrm{\alloy{e}}) \cup \resolvant(\textrm{\alloy{iden}}) \\
  \resolvant(k) & := \{(\fortressint)\} \tag*{for any integer $k$}  \\
  \resolvant(\textrm{\alloy{e1 X e2}}) & := \{(\fortressint)\} \tag*{for any binary integer operator \alloy{X}}  \\
  \resolvant(\textrm{\alloy{#e}}) & := \{(\fortressint)\} \\
  \resolvant(\textrm{\alloy{sum x: e, ... | ...}}) & := \{(\fortressint)\} \\
  \resolvant(\textrm{\alloy{univ}}) & := \{(S) : S \in \mathcal{S}\} \\
  \resolvant(\textrm{\alloy{iden}}) & := \{(S, S) : S \in \mathcal{S}\} \\
  \resolvant(\textrm{\alloy{none}}) & := \emptyset
\end{align*}
\caption{Definition of $\resolvant$: Alloy\_expression $\longrightarrow$ set(sort\_expression).} %
\label{fig:sort-resolvant}
\end{figure*}

In \fortresslang{}, every expression must have a single sort.
Alloy has a signature hierarchy, which means that one atom can be in multiple signatures. Thus, when traversing an Alloy
expression top-down, its sort in \fortresslang\ is not known without consideration of other signatures. 
In translation, multiple signatures may map to the same sort in
\fortresslang{}.
We describe
how the sorts of an Alloy expression are determined for our translation,
inspired by the bounding type system introduced by Edwards, Jackson, and Torlak \cite{Edwards2004a}, where it was used for typechecking.

A \dfn{sort expression} is a tuple of sorts representing the compound sort of a function.
A \dfn{sort resolvant}, $\resolvant(\textrm{\alloy{e}})$, for an Alloy expression \alloymath{e} is a set of sort expressions
that contains every possible sort expression for any tuple of atoms in the interpretation of $\alloymath{e}$. 
For example, if the set of \fortresslang{} sorts is $\mathcal{S} = \{S_1,\dots,S_k\}$, the smallest sort resolvant of \alloy{iden} is $\{(S_1, S_1), (S_2, S_2), \dots, (S_k, S_k)\}$.
Sort resolvants correspond to the types introduced in \cite{Edwards2004a}.

In Figure~\ref{fig:sort-resolvant}, we adapt the calculation of the bounding type in \cite{Edwards2004a} to determine the sort resolvant of a non-Boolean-valued Alloy expression.
In contrast to \cite{Edwards2004a}, at the leaf level of expressions in the Alloy AST, an Alloy signature
is mapped to a sort in the set of \fortresslang{} sorts, $\mathcal{S}$, 
by a $\sortfn$ function (described below).
The definition of $\resolvant$ is independent of the sort policy.

For example, in $\resolvant(\textrm{\alloy{e1 + e2}})$, the possible sorts depend on the signatures of \alloy{e1} and \mbox{\alloy{e2}.}   $\resolvant(\textrm{\alloy{univ}})$
is the set of all sorts.  As another example, $\resolvant(\textrm{\alloy{e1 <: e2}})$ is equal to the resolvants of \alloy{e2} where the resolvant of \alloy{e1} matches the first sort in a sort expression for \alloy{e2}.  If \alloy{e1} never matches the first sort of \alloy{e2} then the resolvant set would be empty, just as the set itself would be empty in any Alloy instance.
If the resolvant of an expression is empty, then we know this 
expression contains no atoms; \myie\ it is equivalent to \alloy{none} of the appropriate arity.
For bound variables, we use the sort resolvant of the bounding expression.
For let-statements and predicate/function calls, we recurse into the substatement or predicate/function body; for predicate/function parameters, we recurse into the passed argument.

Formally, for every tuple of Alloy atoms $(x_1,\dots,x_n)$ that is in 
some interpretation of the expression \alloymath{e} %
where the atom $x_i$ belongs to the top-level Alloy signature $\alloymath{s}_i$, %
 we have $(\sortfn(\text{\alloy{s}}_1),\dots,\sortfn(\text{\alloy{s}}_n)) \in \resolvant(\alloymath{e})$. 
A sort resolvant for an Alloy expression is called \dfn{definite} if it consists of only a single sort expression. If $\resolvant(\textrm{\alloy{e}})$ is definite, then we can assign exactly one sort to each tuple position in 
\alloy{e}.
In Figure~\ref{fig:sort-resolvant}, where the set union operator ($\qunion$) is used are places where the sort resolvant of an expression can be indefinite.

For the leaf-level Alloy expressions, \portus{} includes two possible sort policies. The first, simple sort policy, known as the \dfn{default sort policy}, is defined as follows:
\[ \sortfn_{\textit{default}}(\text{\alloy{S}}) = \begin{cases}
  \fortressint &\textrm{if } \alloymath{S} = \alloymath{Int} \\
  \fortressuniv &\textrm{otherwise}
\end{cases} \]
where $\fortressint$ is the built-in \fortress\ integer sort and $\fortressuniv$ is a sort meant to match Alloy's universe of all atoms.
Our second sort policy is an optimization to give more specific sorts to Alloy's atoms when possible (Section~\ref{sec:partition-sort-policy}).

We note that $\resolvant(\textrm{\alloy{e}})$ is not necessarily the smallest set of sort expressions possible:
there is no guarantee that for each sort expression $(S_1,\dots,S_n) \in \resolvant(\alloymath{e})$, some interpretation of \alloy{e} actually contains a tuple with sorts $(S_1,\dots,S_n)$.
Thus, $\resolvant(\alloymath{e})$ is an overapproximation.

\subsection{Translating Signatures}
\label{sec:sig}

\begin{figure}[h]
\begin{lstlisting}[language=MyAlloy]
sig A { %*\label{lbl:simplesig}*)
    R: A1, %*\label{lbl:rel}*)
    f: B,
    g: f
}
sig A1 extends A {} %*\label{lbl:subsig}*)
sig A2 in A {} %*\label{lbl:subset-sig}*)
sig B {}
\end{lstlisting}
\caption{Alloy signatures example.}
\label{fig:sig-example}
\end{figure}

\begin{table*}[thb]
\begin{center}
\begin{tabular}{|m{5cm}|m{5.2cm}|m{7cm}|} \hline
Description & Example & Axiom \\ \hline \hline
Subsignatures are subsets of non-abstract parent signature & \begin{lstlisting}[language=MyAlloy,numbers=none,framexleftmargin=0cm,xleftmargin=0cm]
sig A {}
sig A1, A2 extends A {}
\end{lstlisting} &
$\trans{\alloytext{A1 + A2 in A}}$ \\ \hline
Subsignatures exactly cover abstract parent signature & \begin{lstlisting}[language=MyAlloy,numbers=none,framexleftmargin=0cm,xleftmargin=0cm]
abstract sig A {}
sig A1, A2 extends A {}
\end{lstlisting} &
$\trans{\alloytext{A1 + A2 = A}}$ \\ \hline
Subset signature is a subset of parents & \begin{lstlisting}[language=MyAlloy,numbers=none,framexleftmargin=0cm,xleftmargin=0cm]
sig A, B, C {}
sig D in A + B + C {}
\end{lstlisting} &
$\trans{\alloytext{D in A + B + C}}$ \\ \hline
Subsignatures do not overlap & \begin{lstlisting}[language=MyAlloy,numbers=none,framexleftmargin=0cm,xleftmargin=0cm]
sig A {}
sig A1, A2 extends A {}
\end{lstlisting} &
$\qforall x: \sortfn(\alloytext{A}) \st \qnot(\trans{x \in \alloytext{A1}} \qand \trans{x \in \alloytext{A2}})$ \newline
for each pair of sibling subsignatures %
\\ \hline
Top-level signatures mapped to same sort do not overlap & \begin{lstlisting}[language=MyAlloy,numbers=none,framexleftmargin=0cm,xleftmargin=0cm]
sig A, B {}
\end{lstlisting}
where $\sortfn(\alloytext{A}) = \sortfn(\alloytext{B}) = S$ &
$\qforall x: S \st \qnot(\trans{x \in \alloytext{A}} \qand \trans{x \in \alloytext{B}})$ \newline
for each pair of top-level signatures %
\\ \hline
Signature multiplicities & \begin{lstlisting}[language=MyAlloy,numbers=none,framexleftmargin=0cm,xleftmargin=0cm]
one sig A {}   // (1)
lone sig B {}  // (2)
some sig C {}  // (3)
\end{lstlisting} &
(1): $\trans{\alloytext{one A}}$ \newline
(2): $\trans{\alloytext{lone B}}$ \newline
(3): $\trans{\alloytext{some C}}$ \\ \hline
\end{tabular}
\end{center}
\caption{Summary of signature axioms.}
\label{tab:sig-axioms}
\end{table*}

A \dfn{signature} in Alloy introduces a set of atoms. For example, line \ref{lbl:simplesig} in Figure~\ref{fig:sig-example} introduces a signature named \alloy{A}. Since \alloy{A} is declared independently of any other signature, it is called a \dfn{top-level} signature.
Each signature has a \dfn{scope}, which is a nonnegative integer that constrains the size of the set: an \dfn{exact scope} sets the exact size of the set, while a \dfn{nonexact scope} sets an upper bound on the set's size.
For every signature (whether top-level or not), \portus{} introduces an \fortresslang\ \dfn{membership predicate} to describe membership in that signature.
For the signature \alloy{A}, we introduce a predicate $\fname{inA} : \sortfn(\alloytext{A}) \to \fortressbool$, where $\fname{inA}(x) = \top$ means that the atom corresponding to $x$ is in \alloy{A}.
Using predicates allows us to distinguish members of different subsignatures in the same sort and lets us handle non-exact scopes %
and signatures of scope zero.

Alloy supports a signature hierarchy. An Alloy \dfn{subsignature}, declared with \alloy{extends}, forms a subset of a single parent signature that does not intersect with any other subsignatures of the same parent.
An Alloy \dfn{subset signature}, declared with \alloy{in}, forms a subset of one or more parent signatures that may intersect with other subsignatures and subset signatures of the parents.
For example, lines \ref{lbl:subsig} and \ref{lbl:subset-sig} in Figure~\ref{fig:sig-example} declare \alloy{A1} as a subsignature of \alloy{A} and \alloy{A2} as a subset signature of \alloy{A}.

\fortresslang\ does not have a sort hierarchy.
Therefore, to handle the signature hierarchy in Alloy, we introduce additional axioms.
Table \ref{tab:sig-axioms} shows the axioms that are added to handle Alloy's signature hierarchy. 
For example, for every signature \alloy{S1} that is a subsignature of a signature \alloy{S}, we add the axiom $\trans{\alloytext{S1 in S}}$ to the \fortress{} theory.
We rely on our translation function (rather than going directly to \fortresslang) so optimizations can be applied to the axiom.
If two top-level signatures \alloy{A} and \alloy{B} are mapped to the same sort $S$ by the sort policy,
we add an axiom ensuring that no atom is a member of both signatures.

The size of each \fortresslang{} sort $S$ except $\fortressint$ (whose size is determined by a bitwidth) is calculated by summing the scopes, whether they are exact or non-exact, of every top-level Alloy signature \alloy{S} such that $\sortfn(\alloytext{S}) = S$. 
If this sum gives a sort size of zero, we bump it up to one, because \fortress{} sorts must have a size of at least one.
For generality, the basic translation method of \portus{} uses 
\textbf{cardinality-based scope axioms} to describe the scopes of Alloy signatures:
\begin{center}
\begin{tabular}{|l|l|} \hline
Description & Axiom \\ \hline \hline
Signature \alloy{A} of exact scope $c$& 
$\trans{\alloytext{#A =}\;\;c}$ \\ \hline
Signature \alloy{A} has non-exact scope $c$ & 
$\trans{\alloytext{#A <=}\;\;c}$ \\ \hline
\end{tabular} 
\end{center}
\noindent The translation of cardinality is described in Section \ref{sec:formulas}. %
We do not include scope axioms for non-exact scopes equal to the size of the sort. Scope axioms are needed for exact scopes equal to the size of the sort to ensure the membership predicate is true for all atoms in the sort.
The bitwidth of integers in the model must be sufficiently large such that the size of every sort 
can be represented as an integer in the model.
These scope axioms ensure that for a signature of scope size 0, even if the sort size is one, the membership predicate of that signature is always false.

\subsection{Translating Fields}
\label{sec:fields}

A \textit{field} in Alloy introduces a relation between the enclosing signature and one or more signatures, bounded by an expression.
For example, line \ref{lbl:rel} in Figure~\ref{fig:sig-example} introduces a field \alloy{R} which is a subset of \alloy{A->A1}.
In general, for any Alloy expression \alloy{e}, a declaration \alloy{sig A \{ f: e \}} introduces a field \alloy{f}, which is a subset of \alloy{A->e}, where \alloy{->} denotes the cross product operation in Alloy.

In our general translation, \portus{} creates an \fortresslang{} predicate to represent each field.
For a field declared as 
\begin{center}
\alloy{sig A \{ f: set e \}}
\end{center}
and $\resolvant(\alloytext{A->e}) = \{(S_1,\dots,S_n)\}$ (required to be definite),
\portus{} declares an \fortresslang{} predicate
\[ f : S_1 \times S_2 \times \dots \times S_n \to \fortressbool \]
and adds the axioms,
\begin{gather} 
\trans{\alloytext{all x: A | x.f in e}}\\ 
\forall x_1: S_1, \dots, x_n: S_n \st \trans{(x_1,\dots,x_n) \in \alloymath{f}} \qimplies \trans{x_1 \in \alloymath{A}}
\end{gather}
to ensure the first element of every tuple is in the domain \alloy{A}. %
References to previously-defined fields in the same signature may occur in \alloy{e}: \myeg{} in Figure~\ref{fig:sig-example}, the field \alloy{g} is bounded by the previously-defined field \alloy{f}.
\portus{} ensures that such fields are correctly referenced in the form \alloy{x.f}.

\begin{figure}
\begin{lstlisting}[language=MyAlloy]
sig A {
    f1: one B,
    f2: lone B,
    f3: some B,
    f4: set B,
    f5: B->set C, %*\label{lbl:arrow-mult}*)
    f6: B one->one C %*\label{lbl:arrow-mult-bijection}*)
}
sig B, C {}
\end{lstlisting}
\caption{Examples of Alloy fields.}
\label{fig:field-mult}
\end{figure}

Alloy fields may have multiplicities associated with them. For example, in Figure~\ref{fig:field-mult}, \alloy{f1} maps each atom in \alloy{A} to exactly one atom in \alloy{B} (the default), \alloy{f2} maps each \alloy{A} to at most one \alloy{B}, \alloy{f3} to at least one \alloy{B}, and \alloy{f4} to any number of atoms of \alloy{B}.
Multiplicities may also occur in expressions formed with the product operator (\alloy{->}): \myeg\ for each atom \alloy{a} in \alloy{A}, line \ref{lbl:arrow-mult} declares that \alloy{a.f5} maps each \alloy{B} to a set of atoms of \alloy{C}, and line \ref{lbl:arrow-mult-bijection} declares that \alloy{a.f6} is a bijection from \alloy{B} to \alloy{C}.

All functions/predicates in F-MSFOL must be total.
\portus{} ignores multiplicity of fields when computing the sort resolvant and declaring the predicate.
For any \mbox{non-\alloy{set}} multiplicity $\alloytext{M} \in \{ \alloytext{one}, \alloytext{lone}, \alloytext{some} \}$ and a field declared as
\alloy{sig A \{ f: M e \}}, \portus{} generates the following axiom (replacing (1)): 
\[ \trans{\alloytext{all x: A | (M x.f) and (x.f in e)}} \]
thereby converting the Alloy multiplicity constraint into an Alloy formula.
When \alloy{e} is an arrow expression containing multiplicities, such as on lines \ref{lbl:arrow-mult} and \ref{lbl:arrow-mult-bijection} of Figure~\ref{fig:field-mult}, the formula \alloy{x.f in e} is an example of what Jackson refers to as a declaration formula \cite{Jackson2012}. The translation of declaration formulas is described in Section \ref{sec:formulas}.

\subsection{Translating Formulas}
\label{sec:formulas}

We translate an Alloy formula to \fortresslang{} by traversing it in a top-down order.
We divide our presentation into sections on similar operators.
We regularly use a technique that we call \dfn{short-circuiting}: if an Alloy formula to be translated relies on a subformula \alloy{e}, but $\resolvant(\alloymath{e}) = \emptyset$, then we know that \alloy{e} statically resolves to \alloy{none}, which permits a simplified translation of the containing formula.
In the language of the bounding type system of Edwards, Jackson, and Torlak \cite{Edwards2004a}, we short-circuit when we detect an empty type, or in some cases, an irrelevancy error.

\begin{figure}
\begin{center}
\begin{align*}
    \trans{\alloymath{not f}} &:= \qnot \trans{\alloymath{f}} \\
    \trans{\alloymath{f1 and f2}} &:= \trans{\alloymath{f1}} \qand \trans{\alloymath{f2}} \\
    \trans{\alloymath{f1 or f2}} &:= \trans{\alloymath{f1}} \qor \trans{\alloymath{f2}} \\
    \trans{\alloymath{f1 => f2}} &:= \trans{\alloymath{f1}} \qimplies \trans{\alloymath{f2}} \\
    \trans{\alloymath{f1 <=> f2}} &:= \trans{\alloymath{f1}} \qiff \trans{\alloymath{f2}} \\
    \trans{\alloymath{f => f1 else f2}} &:= \qite{\trans{\alloymath{f}}}{\trans{\alloymath{f1}}}{\trans{\alloymath{f2}}}
\end{align*}
\caption{Translation of logical operators.}
\label{fig:trans-logical-ops}
\end{center}
\end{figure}

\textbf{Logical operators.} Figure~\ref{fig:trans-logical-ops} shows the translation of the simple logical operators, where \alloy{f}, \alloy{f1}, and \alloy{f2} are Alloy formulas.
The notation $\qite{\mathit{cond}}{\mathit{if\_t}}{\mathit{if\_f}}$ denotes the \fortresslang{} if-then-else operator.

\begin{figure*}
\begin{center}
\begin{align*}
    \trans{(x_1,\dots,x_n) \in \alloymath{\{y1: e1, ..., yn: en | f\}}} & := \trans{x_1 \in \alloymath{e1}} \qand \dots \qand \trans{x_n \in \alloymath{en}} \qand \trans{\alloymath{f}}\\
    \trans{(x_1,\dots,x_n) \in \alloymath{e1 + e2}} &:= \trans{(x_1,\dots,x_n) \in \alloymath{e1}} \qor \trans{(x_1,\dots,x_n) \in \alloymath{e2}} \\
    \trans{(x_1,\dots,x_n) \in \alloymath{e1 & e2}} &:= \trans{(x_1,\dots,x_n) \in \alloymath{e1}} \qand \trans{(x_1,\dots,x_n) \in \alloymath{e2}} \\
    \trans{(x_1,\dots,x_n) \in \alloymath{e1 - e2}} &:= \trans{(x_1,\dots,x_n) \in \alloymath{e1}} \qand \qnot \trans{(x_1,\dots,x_n) \in \alloymath{e2}} \\
    \trans{(x_1,x_2) \in \texttt{\textasciitilde}\alloymath{e}} &:= \trans{(x_2,x_1) \in \alloymath{e}} \\
    \trans{(x_1,\dots,x_n) \in \alloymath{e1 <: e2}} &:= \trans{x_1 \in \alloymath{e1}} \qand \trans{(x_1,\dots,x_n) \in \alloymath{e2}} \\
    \trans{(x_1,\dots,x_n) \in \alloymath{e1 :> e2}} &:= \trans{x_n \in \alloymath{e2}} \qand \trans{(x_1,\dots,x_n) \in \alloymath{e1}} \\
    \trans{(x_1,\dots,x_n) \in \alloymath{e1 -> e2}} &:= \trans{(x_1,\dots,x_m) \in \alloymath{e1}} \qand \trans{(x_{m+1},\dots,x_{n}) \in \alloymath{e2}} \\
    &\qquad\text{where $\arity{\alloymath{e1}} = m$ and $\arity{\alloymath{e2}} = n-m$} \\
    \trans{(x_1,\dots,x_n) \in \alloymath{f => e1 else e2}} &:= \qite{\trans{\alloymath{f}}}{\trans{(x_1,\dots,x_n) \in \alloymath{e1}}}{\trans{(x_1,\dots,x_n) \in \alloymath{e2}}}\\
    \trans{(x_1, \dots, x_n) \in \alloymath{e1.e2}} & := 
     \exists y: S \st \trans{(x_1,\dots,x_m,y) \in \alloymath{e1}} \qand \trans{(y,x_{m+1},\dots,x_n) \in \alloymath{e2}}\\
     & \qquad \text{where $\arity{\alloymath{e1}} = m+1$ and $\arity{\alloymath{e2}} = n-m+1$}\\
    \trans{(x_1,\dots,x_n) \in \alloymath{e1 ++ e2}} & := \trans{(x_1,\dots,x_n) \in \alloymath{e2}} \lor (\trans{(x_1,\dots,x_n) \in \alloymath{e1}} \land {} \\
    &\qquad\lnot(\exists x'_2: S_2,\dots,x'_n: S_n \st \trans{(x_1,x'_2,\dots,x'_n) \in \alloymath{e2}}))\\
    \trans{\alloymath{e1 in e2}} & := \qforall x_1: S_1, \dots, x_n: S_n \st \trans{(x_1,\dots,x_n) \in \alloymath{e1}} \qimplies \trans{(x_1,\dots,x_n) \in \alloymath{e2}} \\
    \trans{\alloymath{e1 = e2}} & := \qforall x_1: S_1, \dots, x_n: S_n \st \trans{(x_1,\dots,x_n) \in \alloymath{e1}} \qiff \trans{(x_1,\dots,x_n) \in \alloymath{e2}}
\end{align*}
\caption{Translation of set operators.}
\label{fig:trans-basic-set-ops}
\end{center}
\end{figure*}

\begin{figure*}[htb]
\begin{center}
\begin{math}
\begin{array}{rcl}

    \trans{\alloymath{all x1: e1, ..., xn: en | f}} & := & 
    \qforall x_1: S_1, \dots, x_n: S_n \st \trans{x_1 \in \alloymath{e1}} \qand \dots \qand \trans{x_n \in \alloymath{en}} \qimplies \trans{\alloymath{f}} \\

    \trans{\alloymath{some x1: e1, ..., xn: en | f}} & := & 
    \qexists x_1: S_1, \dots, x_n: S_n \st \trans{x_1 \in \alloymath{e1}} \qand \dots \qand \trans{x_n \in \alloymath{en}} \qand \trans{\alloymath{f}} \\

    \trans{\alloymath{no x1: e1, ..., xn: en | f}} & := & 
    \qforall x_1: S_1, \dots, x_n: S_n \st \trans{x_1 \in \alloymath{e1}} \qand \dots \qand \trans{x_n \in \alloymath{en}} \qimplies \qnot \trans{\alloymath{f}} \\

    \trans{\alloymath{lone x1: e1, ..., xn: en | f}} &:= &
    \qforall x_1: S_1, \dots, x_n: S_n \st \trans{x_1 \in \alloymath{e1}} \qand \dots \qand \trans{x_n \in \alloymath{en}} \qand \trans{\alloymath{f}} \qimplies\\
    & &\qquad \qforall x'_1: S_1, \dots, x'_n: S_n \st \trans{x'_1 \in \alloymath{e1}} \qand \dots \qand \trans{x'_n \in \alloymath{en}} \qand {} \\  
    && \qquad\trans{\alloymath{f}}[x'_1/x_1,\dots,x'_n/x_n] \qimplies x_1 = x'_1 \qand \dots \qand x_n = x'_n \\

    \trans{\alloymath{one x1: e1, ..., xn: en | f}} & := &
    \qexists x_1: S_1, \dots, x_n: S_n \st \trans{x_1 \in \alloymath{e1}} \qand \dots \qand \trans{x_n \in \alloymath{en}} \qand \trans{\alloymath{f}} \qand {} \\
    &&\qquad\qforall x'_1: S_1, \dots, x'_n: S_n \st \trans{x'_1 \in \alloymath{e1}} \qand \dots \qand \trans{x'_n \in \alloymath{en}} \qand {} \\
    &&\qquad \trans{\alloymath{f}}[x'_1/x_1,\dots,x'_n/x_n] \qimplies x_1 = x'_1 \qand \dots \qand x_n = x'_n

\end{array}
\end{math}
\caption{Translation of quantifiers where $\resolvant(\alloymath{ei}) = \{(S_i)\}$ for each $i$.}
\label{fig:trans-quant}
\end{center}
\end{figure*}

\textbf{Integer operators.}
\portus{} translates integer-valued Alloy expressions and integer literals directly to \fortresslang{} integer expressions and literals.

\textbf{Set operators.}
Alloy expressions represent sets, which cannot be directly represented in \fortresslang{}. Therefore, instead of translating an Alloy expression \alloy{e} directly, we translate $(x_1,\dots,x_n) \in \alloymath{e}$, where $x_1,\dots,x_n$ are \fortresslang{} scalars.
Figure~\ref{fig:trans-basic-set-ops} shows the translation of most of the Alloy set operators, where \alloy{e}, \alloy{e1}, and \alloy{e2} are Alloy expressions and \alloy{f} is an Alloy formula.
When translating an Alloy expression $\trans{\alloymath{g}}$, 
a context is maintained to map the variables in \alloymath{g} 
to their \fortresslang{} variables.
Anywhere that sorts are referenced in the translation, the relevant Alloy expression must have a definite sort (such as where quantification is introduced).

There are a number of opportunities for short-circuiting in the
translation of set operators.
For example, if the resolvants of either \alloymath{e1} or 
\alloymath{e2} are empty in a join operation  
then the expression is equivalent to \alloymath{none}, so \mbox{\alloy{e1.e2}} statically resolves to \alloy{none} and \portus{} short-circuits the translation to $\bot$.
Additionally for join, if there is no common sorts in the last tuple of the resolvant of \alloymath{e1} intersected with the first tuple of the resolvant of \alloymath{e2}, then there are no tuples in the join of these
expressions, and \portus{} can short-circuit to $\bot$.

\textbf{Quantification.}
\portus{} translates quantified expressions 
as described in Figure~\ref{fig:trans-quant}.
For first-order quantifiers, we require that each expression \alloy{ei} has a definite sort so that we can assign a single \fortresslang{} sort to each quantified variable.
The syntax $f[y/x]$ denotes $f$ with the variable $y$ substituted for free instances of the variable $x$.
The translation for second-order quantifiers proceeds similarly, but the sorts $S_i$ in Figure~\ref{fig:trans-quant} are more complex expressions; limitations on translating second-order quantifiers are described in Section~\ref{sec:unimplemented}.
There are opportunities here for short-circuiting.
If any of the sort resolvants are empty, then we know that \alloy{ei} statically resolves to the empty set and the quantification is over the empty set, so we can short-circuit: for $\alloymath{Q} \in \{\alloymath{all}, \alloymath{lone}, \alloymath{no}\}$ the result is $\top$, and for $\alloymath{Q} \in \{\alloymath{one},\alloymath{some}\}$ the result is $\bot$.

For the \alloy{all}, \alloy{some}, and \alloy{no} quantifiers, if a bounding expression \alloy{ei} has multiple possible sorts in its sort resolvant, we split the quantified expression into multiple quantified formulas joined together by \alloy{and} or \alloy{or} before translating.
Each new quantified expression has a copy of \alloy{ei} restricted to one sort.
For example, \alloy{all x: A + Int | f} is split into \alloy{(all x: A | f) and (all x: Int | f)} because the sort policy assigns \alloy{A} and \alloy{Int} to different sorts.
This splitting scheme allows us to translate more quantification expressions and is advantageous for the optimization in Section~\ref{sec:partition-sort-policy}.

Quantifiers may also be applied to expressions (not just variables). The Alloy formula \alloy{Q e}
means the quantification \alloy{Q} is true for the number of elements in the set described by the expression \alloy{e}, \myeg\ \alloy{some A1} means that \alloy{A1} is not empty.  
In this form of quantification, \alloy{e} may be an expression of arbitrary arity and thus must be decomposed into variables each of definite sort, and then the expression is translated as above.
In Alloy, the keyword \alloy{disj} may be used after a quantifier to specify disjointness: \myeg{} \alloy{all disj x, y: e | ...} quantifies over all values of \alloy{x} and \alloy{y} that are not equal \cite{Jackson2012}. %
\portus{} supports this construct by adding a conjunct to the translation that $x$ and $y$ must be distinct.

\textbf{Declaration formulas.}
Alloy allows multiplicities within arrow expressions on the right-hand side of \alloy{in} expressions. 
For example, \alloy{f in A one->one B} expresses that \alloy{f} is a bijection between \alloy{A} and \alloy{B}. 
For any Alloy expressions \alloy{x}, \alloy{e1}, and \alloy{e2}, where $\alloymath{M},\alloymath{N} \in \{\alloymath{one,lone,some,set}\}$, 
$\trans{\alloymath{x in e1 M->N e2}}$ is a conjunction of the following axioms:
\begin{itemize}
    \item $\trans{\alloymath{x in e1->e2}}$; %
    \item if \alloy{N} is not \alloy{set}, $\trans{\alloymath{all a: e1 | N a.x}}$;
    \item if \alloy{M} is not \alloy{set}, $\trans{\alloymath{all b: e2 | M x.b}}$;
    \item if \alloy{e1} is itself an arrow expression containing multiplicities, $\trans{\alloymath{all b: e2 | x.b in e1}}$;
    \item if \alloy{e2} is itself an arrow expression containing multiplicities, $\trans{\alloymath{all a: e1 | a.x in e2}}$.
\end{itemize}

\textbf{Transitive closure.}
The closure operators in \fortresslang{} act on a binary predicate, but the Alloy closure operators may act on any binary expression.
 \portus{} introduces an auxiliary predicate definition equal to 
 that binary expression and then takes the closure of the auxiliary predicate.
Consider the translation $\trans{(x_1, x_2) \in \alloymath{^e}}$ for an arbitrary Alloy expression \alloy{e}.
\alloy{e} may include variables \mbox{\alloy{a1,...,an}} that are bound outside of \alloy{e}
and are already mapped to \fortresslang{} variables $a_1,\dots,a_n$
with respective sorts $S_1,\dots,S_n$.
\portus{} introduces a fresh predicate definition $\aux : S \times S \times S_1 \times \dots \times S_n \to \fortressbool$, to translate a closure as:
\begin{center}
\begin{tabular}{l}
$\aux(x'_1,x'_2,a'_1,\dots,a'_n) := \trans{(x'_1,x'_2) \in \alloymath{e}}$\\
$\trans{(x_1,x_2) \in \alloymath{^e}} := Closure(\aux,x_1,x_2,a_1,\dots,a_n)$
\end{tabular}
\end{center}
where \alloy{a1,...,an} are mapped to $a'_1,\dots,a'_n$ when translating $\trans{(x'_1,x'_2) \in \alloymath{e}}$.
The translation of reflexive closure is similar.  The use of definitions for transitive
closure (rather than axioms) results in better performance in the \zthree\ SMT solver~\cite{DeMoura2008}.

\textbf{Cardinality and \alloy{sum}.}
The Alloy set cardinality operator \mbox{\alloy{#e}} returns the number of elements contained in the set represented by the expression \alloymath{e} as an integer value.
\portus{} takes the novel approach of representing cardinality in \fortresslang{} by summing the value one for each tuple of domain elements of the sorts that could be in the set expression.
For instance, if  $\resolvant(\alloymath{e}) = \{(S_1,S_2)\}$, and $S_1 = \{d_{1,1},d_{1,2},d_{1,3}\}$ (scope size three) and $S_2 = \{d_{2,1},d_{2,2}\}$ (scope size two), \alloymath{#e} translates as:
\begin{align*}
    \trans{\alloymath{#e}} :={}&(\qite{\trans{(d_{1,1},d_{2,1}) \in \alloymath{e}}}{1}{0})\\
    {}+{}&(\qite{\trans{(d_{1,1},d_{2,2}) \in \alloymath{e}}}{1}{0})\\
    {}+{}&(\qite{\trans{(d_{1,2},d_{2,1}) \in \alloymath{e}}}{1}{0})\\
    {}+{}&(\qite{\trans{(d_{1,2},d_{2,2}) \in \alloymath{e}}}{1}{0})\\
    {}+{}&(\qite{\trans{(d_{1,3},d_{2,1}) \in \alloymath{e}}}{1}{0})\\
    {}+{}&(\qite{\trans{(d_{1,3},d_{2,2}) \in \alloymath{e}}}{1}{0})
\end{align*}
The scope for integers supplied by the modeller must be large enough to represent the maximum sum.
The use of domain elements within a theory would normally limit the symmetry breaking that can be performed on these sorts by \fortress~\cite{Poremba2023}. However, the domain elements in the above translation are used symmetrically, so we have extended \fortress's symmetry breaking algorithm to recognize this particular pattern of use of domain elements to allow symmetry breaking to continue to occur on the sorts used in a cardinality operation.

The Alloy integer expression \alloy{sum x1: e1, ..., xn: en | f} represents the integer expression \alloy{f} summed over all values of \alloy{x1}, $\dots$, \alloy{xn}. \portus{} translates \alloy{sum} expressions similar to the above translation for cardinality, expanding over each possible value of the quantified variables and summing $\trans{\alloymath{f}}$.

\textbf{Ordering module.}
Alloy's well-used ordering module takes a signature as a parameter and adds axioms to create a total order on the atoms in the signature via relations \alloy{first} and \alloy{next}.
As is done in \kodkod, since the atoms of an Alloy signature are interchangeable, we can add symmetry breaking formulas to require any instance to use a specific ordering of the atoms in the signature.
In this case, \portus{} can add more symmetry breaking constraints than \fortress{} could determine using its general techniques because we know that \alloy{next} defines the total order.
Instances with other orders of the elements are all isomorphic to the order chosen, and therefore need not be explored.

For an ordered set \alloy{A} with exact\footnote{Alloy requires ordered signatures to have exact scopes \cite{Jackson2012}.} scope $k$ and a sort $S = \sortfn(\alloymath{A})$ where $S = \{a_1,\dots,a_k, \dots \} $ are domain elements,
\portus{} assigns $k$ of the domain elements in $S$ to \alloy{A}, picks an arbitrary order of them, and adds the following definition and 
axioms:
\begin{align*}
    \mathord{next}(x):={}&\qite{x=a_1}{a_2}{ \\
    &\qite{x=a_2}{a_3}{ \\
    &\dots \\
    &\qite{x=a_{k-2}}{a_{k-1}}{a_k}}}
\end{align*}
\[ \trans{x \in \alloymath{first}} := x = a_1 \]
\[ \trans{(x_1,x_2) \in \alloymath{next}} := \trans{x_1 \in \alloymath{A}} \qand x_1 \ne a_k \qand \mathord{next}(x_1) = x_2 \]
$\mathord{next}(x)$ is not meaningful when $x$ is the last element $a_k$ or a domain element in $S$ other than those assigned to \alloy{A}.
Therefore, these values are assigned arbitrarily in the above definition.
This assignment of sets of domain elements to signatures is done in a manner that respects the signature hierarchy.
The guard condition $\trans{x_1 \in \alloymath{A}} \qand x_1 \ne a_k$ ensures that the call to $\mathord{next}$ only affects the truth value of the formula when its argument is in the domain where it is meaningful. 
As a slight optimization, we also add a similar function definition $\mathord{prev} : S \to S$ for \alloy{prev} (defined in \texttt{ordering.als} as \texttt{\textasciitilde}\alloy{next}) if it is used.

\begin{figure}[htb]
\begin{center}
\begin{align*}
    \trans{x \in \alloymath{A}} &:= inA(x) && \text{where \alloy{A} is a signature} \\
    \trans{(x_1,\dots,x_n) \in \alloymath{f}} &:= f(x_1,\dots,x_n) && \text{where \alloy{f} is a field} \\
    \trans{x \in \alloymath{v}} &:= x = v && \text{where \alloy{v} is a bound}\\
    & & & \text{variable mapped to $v$} \\
    \trans{x \in \alloymath{univ}} &:= \top \\
    \trans{x \in \alloymath{none}} &:= \bot \\
    \trans{(x_1, x_2) \in \alloymath{iden}} &:= x_1 = x_2 \\
    \trans{\alloymath{let y=e | f}} &:= \trans{\alloymath{f}} && \text{where \alloy{y} is substituted} \\
    & & & \text{with \alloy{e} in $\trans{\alloymath{f}}$}
\end{align*}
\end{center}
\caption{Translation of leaf formulas.}
\label{fig:trans-leaf}
\end{figure}

\textbf{Leaf formulas.}
All translations reduce to combinations of $(x_1,\dots,x_n) \in \alloymath{X}$, where \alloy{X} is a signature, field, variable, or constant with trivial translations as shown in Figure~\ref{fig:trans-leaf}.

\textbf{Functions and predicates.}
Consider an Alloy predicate \alloy{pred p[y1: e1, ..., yn: en] \{ f \}}.
A naïve \fortresslang{} translation would translate calls to \alloy{p} as $\trans{\alloymath{p[z1,...,zn]}} := \trans{\alloymath{f}}$, where the parameter names are mapped to the arguments in the translation of \alloy{f} as appropriate.
However, this would result in repeating $\trans{\alloymath{f}}$ every time \alloy{p} is called.
We cannot create an \fortresslang{} function definition $p(y_1,\dots,y_n)$ exactly representing \alloy{p} because arbitrary Alloy expressions might be passed as arguments to \alloy{p}, but a function definition can only take single atoms as parameters: for instance, there is no simple way to translate \alloy{p[A, ..., A]} in terms of $p(y_1,\dots,y_n)$, where \alloy{A} is a (non-\alloy{one}) signature.

Instead, to maintain the structure of the model, \portus{} creates \fortresslang{} function definitions for each function or predicate \textit{call}.
When translating $\trans{\alloymath{p[z1,...,zn]}}$, \portus{} adds a fresh function definition
$p(a_1,\dots,a_m) := \trans{\alloymath{f}}$
where $a_1,\dots,a_m$ are the free variables that appear in the translation of the arguments \alloy{z1,...,zn}
and the parameter names \alloy{y1,...,yn} are again mapped to \alloy{z1,...,zn} in the translation of \alloy{f}.
\portus{} then translates $\trans{\alloymath{p[z1,...,zn]}} := p(a_1,...,a_m)$.
Our method is similar for Alloy function calls.
Multiple definitions may be created when the same predicate/\allowbreak{}function is called with different arguments;
for example, \alloy{p[x]} and \alloy{p[x+y]} will be translated using different definitions.
These predicate definitions are reused when possible. 
We reuse definitions whenever the call is 
alpha-equivalent to a previously-translated call with respect to the free variables in the arguments.
In practice, we find that in translating real-world models, many definitions are reused.

For example, consider an Alloy predicate \alloy{pred nonempty[x: set A] \{ some x \}}, where \alloy{A} is a signature.
Suppose \alloy{y: A} and \alloy{z: A} are bound variables in the current scope corresponding to \fortresslang{} variables $y$ and $z$.
When \portus{} translates the call \alloy{nonempty[y]}, it creates a predicate definition $\mathit{nonempty}_1(y) := \trans{\alloymath{some y}}$ because \alloy{y} is the only free variable in \mbox{\alloy{nonempty[y]}.}
Then we translate $\trans{\alloymath{nonempty[y]}} = \mathit{nonempty}_1(y)$.
If we then translate \alloy{nonempty[z]}, the definition is reused: $\trans{\alloymath{nonempty[z]}} = \mathit{nonempty}_1(z)$.
But \alloy{nonempty[y+z]} is not alpha-equivalent to \alloy{nonempty[y]}, so to translate it, \portus{} creates a new predicate definition $\mathit{nonempty}_2(y,z) := \trans{\alloymath{some (y+z)}}$
and translates $\trans{\alloymath{nonempty[y+z]}} = \mathit{nonempty}_2(y,z)$.

\subsection{Unsupported Alloy features}
\label{sec:unimplemented}

While Alloy is a language built around clean, understandable
concepts, it has some remarkably subtle language features.
We do not yet translate some less-used Alloy features, such as the bitshift operators and recursion in functions and predicates (\kodkod{} imposes a limit on recursion, which we do not yet do in \portus{}).
We also do not support the temporal logic operators introduced in Alloy 6.
\portus{} does not support ordering subset signatures (those declared with \alloy{in}), which are rarely used \cite{Eid2023}.
\portus{} also does not support ordering both a subsignature and its parent or other ancestor signature since choosing an arbitrary order for both signature's atoms as described above would result in a fixed relationship between the orderings, violating the independence of symmetry breaking constraints.

We do not support existential second-order quantification in places where definitions are required in our translation.  For example, 
\portus{} cannot translate \alloy{some r: A->A | x->y in ^r} because taking the transitive closure of the second-order variable \alloy{r} requires creating an auxiliary definition for the formula $(x_1,x_2) \in \alloymath{r}$, which would need to take the second-order \fortresslang{} variable corresponding to \alloy{r} as a parameter.
\fortresslang{} does not allow passing second-order variables as parameters to function definitions.

In several places in the translation (\myeg{} field bounding expressions and \alloy{in} and \alloy{=} formulas), \portus{} requires Alloy expressions to have definite sort resolvants.  
In our optimizations for sorts described in Section~\ref{sec:partition-sort-policy}, we support signature merging when needed in resolving sorts. Therefore, the limitation to definite sorts in our optimizations
only affects expressions that mix integers and other atoms in places where definite sort resolvants are required.
Our sort policies always map integers  to \fortress's built-in integer sort.
For example, \portus{} cannot translate the field declarations \alloy{sig A \{ f: B + Int \}} or \alloy{sig A \{ f: univ \}} because the expressions \alloy{B + Int} and \alloy{univ} mix integers and other atoms.
This limitation is common in related work (\myeg~\cite{Mohamed2019}, \cite{Arkoudas2003} explicitly mention this limitation and other work does not discuss signature merging).

\newcommand*{\dom}{\mathsf{dom}}
\newcommand*{\abs}[1]{\left|#1\right|}

\section{Optimizations}
\label{sec:opt}
Our general translation and default sort policy do 
not take advantage of any of the special features of F-MSFOL.  For 
example, all signatures are grouped together in one \fortresslang{} sort; few expressions are treated as scalars; and all Alloy fields become predicates with extra axioms.  
Treating all fields as relations over a universal set
is used in \kodkod{} 
to facilitate 
the conversion to propositional logic.  
However, the decision procedures of SMT solvers
work with functions, scalars, and multiple sorts.  In this section, we discuss novel methods for optimizing our general translation through a more precise sort policy, extensive recognition of scalars and functions, and an alternative method for representing scope limitations.
All of our methods are designed to produce performance boosts because \fortress{} symmetry breaking and
SMT decision procedures can take advantage of the structure found in functions and sorts to limit the search space.
Where we refer to statistics regarding the occurrence of constructs in 
Alloy models in this section, we are referencing an analysis of 1652 Alloy models
scraped from the web by Eid and Day~\cite{Eid2023}.
Our \portus{} implementation uses the chain-of-responsibility design pattern~\cite{Gamma1994} for implementing optimizations so the applicability of each optimization is considered prior to defaulting to the general translation.

\subsection{Partition Sort Policy and Membership Predicates}
\label{sec:partition-sort-policy}

The default sort policy used in the general translation groups all non-integer signatures together in one sort.  However,
using more sorts can reduce the size of the possible 
interpretations of a problem; \myeg\ rather than an uninterpreted 
function having all possible values within the space \mbox{\alloy{univ 
-> univ},} it is limited to values in the space \alloy{A -> B}.
Furthermore, more sorts improve \fortress's ability to do symmetry breaking (which is done per sort), and thus decrease the search space, leading to better performance.  Finally, if a quantified variable draws from a more limited set of domain values, quantifier expansion in the \fortress{} FMF method will result in a decreased formula size.

Our \dfn{partition sort policy} assigns each top-level signature to a 
distinct sort when possible. Eid and Day~\cite{Eid2023} observed that 22.6\% of all signatures in their corpus of models are top-level signatures with a model containing four top-level signatures on average, which indicates the potential to partition the set of signatures into multiple sorts.  
However, Alloy expressions may arbitrarily combine different top-level Alloy signatures, so the mapping from signatures to sorts cannot be determined only from the Alloy signature hierarchy.
For example, to translate \alloy{sig S \{ f: A + B \}} %
\portus{} must assign a definite sort to the range of the field \alloy{f}; therefore \alloy{A+B} must have a single sort and the partition sort policy must map \alloy{A} and \alloy{B} to the same sort.

\textbf{Signature Merging.} For this optimization, 
\portus{} computes a partition of the set 
of signatures by traversing the Alloy expressions and merging two signatures into a single partition if a definite sort is needed for that combination of signatures. 
Not all instances of \alloy{A+B} will result in merging the sorts for \alloy{A} and \alloy{B}.
For example, the formula \alloy{A in A+B} will be translated as $\qforall x: \sortfn(\alloymath{A}) \st \trans{x \in \alloymath{A}} \qimplies \trans{x \in \alloymath{A}} \qor \trans{x \in \alloymath{B}}$.
Short-circuiting ensures that $\trans{x \in \alloymath{B}}$ is translated to $\bot$ since the sorts of $x$ and \alloy{B} do not match, leading to an efficient translation result.
Additionally, the quantifier-splitting scheme discussed in Section~\ref{sec:formulas} is applied before computing the partition: hence \alloy{all x: A+B | f} is split into \alloy{(all x: A | f) and (all x: B | f)}, again resulting in fewer sorts having to be merged.
Note that $\fortressint$ cannot be merged with other sorts because it is built-in.

A sort policy matches a signature to exactly one sort.
If a top-level signature has an exact scope equal to the size of its sort (\myie{} it is the only top-level signature in its sort), then we are able to remove the use of the membership predicate for that signature introduced in Section~\ref{sec:sig}.
In this case, the membership predicate would always be true since the sort's size is the same as that of its associated top-level signature.
With more sorts from the partition sort policy, there are more times when membership predicates are not needed.
Additionally, with more sorts, there are more opportunities to remove the scope axioms when a signature has an exact scope equal to the size of the sort.
$\trans{x \in \alloymath{A}}$ where \alloy{A} is such a signature is translated to $\top$ when the variable $x$ is statically determined to be in the sort $\sortfn(\alloymath{A})$ and short-circuited to $\bot$ when $x$ is not in $\sortfn(\alloymath{A})$.

\subsection{Scalar Optimizations}
\label{sec:scalar-opt}

In the general translation,
translating an Alloy expression \alloy{e} is done by translating $(x_1,\dots,x_n) \in \alloymath{e}$ as a formula.  The optimizations we describe in this 
section are based on recognizing combinations of Alloy expressions that represent a single value (a \textbf{scalar}) and then translating those expressions directly into \fortresslang{} scalar values.  By translating to scalars, we can turn predicates into functions and eliminate quantifiers found in the general translation, thus reducing the quantifier expansion in \fortress{}'s FMF method and allowing SMT decision procedures to process functions of reduced arity.

We call an Alloy expression \alloy{e} that contains tuples of arity $n+1$ (where $n$ can be zero) a \dfn{scalar expression} if there exists an \fortresslang{} expression $e$ and an \fortresslang{} formula $\fname{guard}$ (both of arity $n$),\footnote{
Technically, $e$ and $\fname{guard}$ are meta-level functions that create \fortresslang{} formulas.
} such that,\\
$\hspace*{1cm}\trans{(x_1,\dots,x_n,y) \in \alloymath{e}} \equiv $\\
\hspace*{2cm}$\fname{guard}(x_1,\dots,x_n) \qand y = e(x_1,\dots,x_n)$\\
where $\equiv$ denotes semantic equivalence.  
\alloy{e} itself can be represented, 
\myie\ \textbf{cast}, as a scalar  in \fortresslang{}.
$\fname{guard}$ is a formula that is true if and only if \alloy{e} is nonempty, which means the \fortresslang{} expression $e$ applied to arguments represents the single atom corresponding to \alloy{e}.  

If $n$ is zero, the definition of a scalar expression simplifies to:
\[ \trans{y \in \alloymath{e}} \equiv \fname{guard} \qand y = e \]
and $\trans{y \in \alloymath{e}} := y = e$ when $\fname{guard}$ is $\top$.
For example, suppose \alloy{x} is a bound variable from the quantifier in \alloy{all x: S | ...}, which implicitly means \alloy{one x}, \myie\ \alloy{x} represents one value.  If \portus{} has translated \alloy{x} to the \fortresslang{} variable $x$ then 
$\trans{y \in \alloymath{x}} := y = x$.

For generality in recognizing that combinations of scalar expressions are also scalar expressions,
if \alloy{e} is a scalar expression, we create an abstraction for the translation
process called $\casttoscalar$, which is a partial function that is defined as
$$\casttoscalar(\alloymath{e}) = (e, \fname{guard})$$
where 
$e$ and $\fname{guard}$ are both functions of 
the same arity as \alloy{e} (and same sorts) that create \fortresslang{} functions. 
If $\casttoscalar(\alloymath{e})$ has a value, it means
that the Alloy expression \alloy{e} contains at most one atom.
In this section, we present techniques for recognizing scalar expressions, recognizing combinations of scalar expressions, and several optimizations that take advantage of scalar expressions.

\subsubsection{Recognizing Scalars}
\label{sec:scalar-opt-recognizing}

In this section, we describe patterns where we can 
recognize that an Alloy expression is a scalar expression
to result in a definition of $\casttoscalar(\alloymath{e})$ for Alloy 
expression \alloy{e}.  

\begin{figure}[htb]
\begin{center}
\begin{math}
\begin{array}{rcl}
    \casttoscalar(\alloymath{v}) &:= &(v, \top) \\
    \multicolumn{3}{r}{\text{where \alloy{v} is a quantified variable mapped to $v$}} \\
    \casttoscalar(\alloymath{i}) &:= &(\trans{\alloymath{i}}, \top) \\
    \multicolumn{3}{r}{\text{where \alloy{i} is an integer expression}} \\
    \casttoscalar(\alloymath{first}) &:= &(a_1, \top) \\
    \casttoscalar(\alloymath{next}) &:= & \\
    \multicolumn{3}{r}{\qquad\qquad(\lambda x \st \mathord{next}(x), \lambda x \st \trans{x \in \alloymath{A}} \qand x \ne a_k)} \\
    \casttoscalar(\alloymath{prev}) &:= &\\
    \multicolumn{3}{r}{(\lambda x \st \mathord{prev}(x), \lambda x \st \trans{x \in \alloymath{A}} \qand x \ne a_1)} \\
    \multicolumn{3}{r}{\text{where the ordered sig \alloy{A} is assigned}}\\ 
    \multicolumn{3}{r}{\text{domain elements $a_1,\dots,a_k$}} \\
\end{array}
\end{math}
\end{center}
\caption{Basic scalar expressions recognized by \portus{}.}
\label{fig:casttoscalar-basic}
\end{figure}

\textbf{Constants and variables.}  Alloy expressions that are quantified variables, integer expressions, or constants (including built-in functions),
can immediately be recognized as scalars as is shown with 
the definitions of $\casttoscalar$ in Figure~\ref{fig:casttoscalar-basic}.

\textbf{Signatures with multiplicity one.}
Alloy signatures declared with the multiplicity \alloy{one}, known as \dfn{one sigs}, can usually be considered scalar expressions.
Eid and Day~\cite{Eid2023} observed that 26.4\% of all signatures in their corpus of models were \alloy{one} sigs.
If \alloy{A} is a \alloy{one} sig, \portus{} assigns \alloy{A} to a specific \fortresslang{} domain element\footnote{Any direct assignment of a domain element to an Alloy expression is a form of symmetry breaking.} $a \in \sortfn(\alloymath{A})$ 
and 
\[ \casttoscalar(\alloymath{A}) := (a, \top) \] where the guard is $\top$ because \alloy{A} is never empty.
Additionally, we do not generate the membership predicate for \alloy{A}.

This optimization cannot be applied if \alloy{A} is a subsignature (at any depth) of any signature that has been ordered. When translating the ordering module as explained in Section~\ref{sec:formulas}, \portus{} chooses an arbitrary domain element order, and assigning \alloy{A} to a particular domain element $a$ in the order would incorrectly fix the position of \alloy{A} in the order.
The optimization is also not applied when \alloy{A} is a subset signature %
because subset signatures are not guaranteed to be disjoint from other signatures and those other signatures could be one sigs or be ordered.
Eid and Day~\cite{Eid2023} found only 0.2\% of one sigs are subset signatures.

\begin{figure}
\begin{lstlisting}[language=MyAlloy]
sig A {
    f: one B,
    g: lone B,
    h: B->one C
}
sig B, C {}
\end{lstlisting}
\caption{Example of functions in Alloy.}
\label{fig:func-opt-example}
\end{figure}

\textbf{Total and Partial Functions.} 
Eid and Day~\cite{Eid2023} found that 53.6\% of fields were total functions (declared with range multiplicity \alloy{one}) and 
12.3\% of fields were partial functions (declared with range multiplicity \alloy{lone})
in their corpus.
For example, in Figure~\ref{fig:func-opt-example}, \alloy{f} is a total function from \alloy{A} to \alloy{B}, \mbox{\alloy{g}} is a partial function from \alloy{A} to \alloy{B}, and \alloy{h} is a binary total function from \alloy{A->B} to \alloy{C}.
Since these are functions, \portus{} can optimize their translation to make them functions in \fortresslang{} rather than relations.

An Alloy total or partial function has the form
$\alloymath{sig S \{ f: e->M R \}}$,
where $\alloymath{M} \in \{\alloymath{one}, \alloymath{lone}\}$
and \alloy{e} is an arbitrary Alloy expression of any arity $n$ ($n$ may be zero).
Instead of generating a predicate for \alloymath{f} (as described in Section~\ref{sec:fields}), 
\portus{} declares a fresh \fortresslang{} function
\[ f: S \times E_1 \times E_2 \times \cdots \times E_n \to R \]
in the theory, where $\alloymath{e}$ is an Alloy expression whose sorts resolve to $E_1 \times E_2 \times \cdots \times E_n$ and $\resolvant(\alloymath{S->e->R}) = \{(S,E_1,E_2,\dots,E_n,R)\}$ (where $\resolvant$ is as defined in Fig.~\ref{fig:sort-resolvant}).
Even when \alloy{f} is a total function, the domain of \alloy{f} may not include all tuples in $S \times E_1 \times \cdots \times E_n$: only the tuples in the Alloy expression \alloy{S->e} are in the domain of \alloy{f}.
We use the syntax $(s,x_1,\dots,x_n) \in \dom(\alloymath{f})$ to mean that $(s,x_1,\dots,x_n)$ is in the domain of \alloy{f} and define its translations.
We add the following axiom to constrain $f$:
\begin{align*}
    &\forall s: S, x_1: E_1, \dots, x_n: E_n \st {} \\
    &\quad \trans{(s,x_1,\dots,x_n) \in \dom(\alloymath{f})} 
    \qimplies \trans{f(s,x_1,\dots,x_n) \in \alloymath{R}}
\end{align*}
In \fortresslang{}, $f$ is a total function, but our translation only cares about its result for values in its domain.  Recall that in Figure~\ref{fig:trans-basic-set-ops}, all expressions include limitations on the domain values being within the translation of the signatures (\myeg\ $\trans{x_1 \in \alloymath{e}}$).

Next we show how to translate the domain expressions.
When \alloy{f} is a total function (declared with \alloy{one}):
\begin{align*}
    \trans{(s,x_1,\dots,x_n) \in \dom(\alloymath{f})} := \trans{(s,x_1,\dots,x_n) \in \alloymath{S->e}}
\end{align*}
If \alloy{f} is declared with \alloy{lone}, the domain of \alloy{f} may be any subset of \alloy{S->e} (rather than being exactly equal to \alloy{S->e}). To capture a
subset of \alloy{S->e}, \portus{} adds a fresh \fortresslang{} predicate
\[ \fname{inDomain}_f : S \times E_1 \times \cdots \times E_n \to \fortressbool \]
to the theory, and we set
\[ \trans{(s,x_1,\dots,x_n) \in \dom(\alloymath{f})} := \fname{inDomain}_f(s,x_1,\dots,x_n) \]
The following axiom makes any tuples that satisfy $\fname{inDomain}_f$ be within \alloy{S->e}:
\begin{align*}
    &\forall s: S, x_1: E_1, \dots, x_n: E_n \st {} \\
    & \quad \fname{inDomain}_f(s,x_1,\dots,x_n) \qimplies \trans{(s,x_1,\dots,x_n) \in \alloymath{S->e}}
\end{align*}
Finally,
for a total or partial function \alloy{f}
declared as above, %
\begin{center}
$\casttoscalar(\alloymath{f}) := (f, \fname{guard}_f)$
$\fname{guard}_f(s,x_1,\dots,x_n) := \trans{(s,x_1,\dots,x_n) \in \dom(\alloymath{f})}$ 
\end{center}

\subsubsection{Combining Scalar Expressions}

Our $\casttoscalar$ abstraction allows us to recognize that
combinations of scalar expressions are also scalar expressions.
We can build rules for compositions of various Alloy expressions.
For Alloy's \textbf{join} operator, if
\begin{center}
\begin{math}
\begin{array}{l} 
\casttoscalar(\alloymath{f1}) := (e_1, \fname{guard}_1)\\
\casttoscalar(\alloymath{f2}) := (e_2, \fname{guard}_2)
\end{array}
\end{math}
\end{center}
the rule is:
\begin{center}
\begin{math}
\begin{array}{l} 
\casttoscalar(\alloymath{f1.f2}) := (e_c, \fname{guard}_c)\\
\text{where}\\
e_c(x_1,\dots,x_n,y_2,\dots,y_m) := e_2(e_1(x_1,\dots,x_n),y_2,\dots,y_m) \\
\fname{guard}_c(x_1,\dots,x_n,y_2,\dots,y_m) := \\
\quad \fname{guard}_1(x_1,\dots,x_n) 
\qand \fname{guard}_2(e_1(x_1,\dots,x_n),y_2,\dots,y_m)
\end{array}
\end{math}
\end{center} 
\newcommand{\xs}{\bar{x}}
\newcommand{\ys}{\bar{y}}

\begin{figure*}
\begin{center}
\begin{math}
\begin{array}{rcl}
    \casttoscalar(\alloymath{e1 \& e2}) &:= &(\lambda \xs \st e_1(\xs), \lambda \xs \st \fname{guard}_1(\xs) \land \trans{(\xs,e_1(\xs)) \in \alloymath{e2}}) \\
    \casttoscalar(\alloymath{e2 \& e1}) &:= &(\lambda \xs \st e_1(\xs), \lambda \xs \st \fname{guard}_1(\xs) \land \trans{(\xs,e_1(\xs)) \in \alloymath{e2}}) \\
    \multicolumn{3}{r}{\text{where}~\casttoscalar(\alloymath{e1}) = (e_1, \fname{guard}_1)} \\
    \casttoscalar(\alloymath{e1 - e2}) &:= &(\lambda \xs \st e_1(\xs), \lambda \xs \st \fname{guard}_1(\xs) \land \neg\trans{(\xs,e_1(\xs)) \in \alloymath{e2}}) \\
    \multicolumn{3}{r}{\text{where}~\casttoscalar(\alloymath{e1}) = (e_1, \fname{guard}_1)} \\
    \casttoscalar(\alloymath{e1 <: e2}) &:= &(\lambda \xs \st e_2(\xs), \lambda \xs \st \fname{guard}_2(\xs) \land \trans{x_1 \in \alloymath{e1}}) \\
    \multicolumn{3}{r}{\text{where}~\casttoscalar(\alloymath{e2}) = (e_2, \fname{guard}_2)} \\
    \casttoscalar(\alloymath{e1 :> e2}) &:= &(\lambda \xs \st e_1(\xs), \lambda \xs \st \fname{guard}_1(\xs) \land \trans{e_1(\xs) \in \alloymath{e2}}) \\
    \multicolumn{3}{r}{\text{where}~\casttoscalar(\alloymath{e1}) = (e_1, \fname{guard}_1)} \\
    \casttoscalar(\alloymath{e1 :> e2}) &:= &(\lambda \xs \st e_2, \lambda \xs \st \fname{guard}_2 \land \trans{(\xs, e_2) \in \alloymath{e1}}) \\
    \multicolumn{3}{r}{\text{where}~\casttoscalar(\alloymath{e2}) = (e_2, \fname{guard}_2)~\text{and \alloy{e2} is unary}} \\
    \casttoscalar(\alloymath{e1 ++ e2}) &:= &(\lambda \xs \st \qite{\fname{guard}_2(\xs)}{e_2(\xs)}{e_1(\xs)}, \lambda \xs \st \fname{guard}_1(\xs) \lor \fname{guard}_2(\xs)) \\
    \multicolumn{3}{r}{\text{where}~\casttoscalar(\alloymath{e1}) = (e_1, \fname{guard}_1)~\text{and}~\casttoscalar(\alloymath{e2}) = (e_2, \fname{guard}_2)~\text{and \alloy{e1} and \alloy{e2} are binary}} \\
    \casttoscalar(\alloymath{e1 -> e2}) &:= &(\lambda (\xs,\ys) \st e_2(\ys), \lambda (\xs,\ys) \st \fname{guard}_2(\ys) \land \trans{\xs \in \alloymath{e1}}) \\
    \multicolumn{3}{r}{\text{where}~\casttoscalar(\alloymath{e2}) = (e_2, \fname{guard}_2)~\text{and \alloy{e2} is unary and \alloy{e1} has a definite sort resolvant}} \\
    \casttoscalar(\texttt{\textasciitilde}\alloymath{(e1 -> e2)}) &:= &(\lambda x \st e_1, \lambda x \st \fname{guard}_1 \land \trans{x \in \alloymath{e2}}) \\
    \multicolumn{3}{r}{\text{where}~\casttoscalar(\alloymath{e1}) = (e_1, \fname{guard}_1)~\text{and \alloy{e1} and \alloy{e2} are unary and have definite sort resolvants}}
\end{array} 
\end{math}
\end{center}
\caption{
Relational expressions recognized as scalars by \portus{}. Here \alloy{e1} has arity $n$, \alloy{e2} has arity $m$, and we abbreviate $\xs = (x_1,\dots,x_k)$ and $\ys = (y_1,\dots,y_\ell)$, where the arities $k$ and $\ell$ are clear from context. %
When not specified, the Alloy expressions \alloy{e1} and \alloy{e2} may be recognized as scalar expressions or not; further optimizations are applied if more scalar expressions are recognized.
}
\label{fig:casttoscalar-relational}
\end{figure*}

\noindent This general rule for join, which recognizes that joins of scalar expressions are also scalar expressions, allows many Alloy expressions to be recognized as scalars.
For example (with applications of $\lambda$-expressions reduced), 
\begin{center}
\begin{math}
\begin{array}{l} 
\text{if}\\
\casttoscalar(\alloymath{f}) := (\lambda y \st f(y), \lambda y \st \fname{inA}(y))\\
\casttoscalar(\alloymath{x}) := (x, \top)\\ 
\text{then} \\
\casttoscalar(\alloymath{x.f}) := (f(x), \top \qand \fname{inA}(x))
\end{array}
\end{math}
\end{center} 
meaning \alloy{x.f} is recognized as a scalar expression by this rule.  Here is another example showing composition of Alloy functions:
\begin{center}
\begin{math}
\begin{array}{l} 
\text{if}\\
\casttoscalar(\alloymath{f}) := (\lambda x \st f(x), \lambda x \st \fname{inA}(x)) \\
\casttoscalar(\alloymath{g}) := (\lambda y \st g(y), \lambda y \st \fname{inB}(y))\\
\text{then} \\
\casttoscalar(\alloymath{f.g}) := \\
\quad (\lambda x \st g(f(x)), \lambda x \st \fname{inA}(x) \qand \fname{inB}(f(x)))
\end{array}
\end{math}
\end{center} 
Another example showing a join expression for a multi-arity Alloy function is:
\begin{center}
\begin{math}
\begin{array}{l} 
\text{if}\\
\casttoscalar(\alloymath{h}) := \\
\quad (\lambda (x, y) \st h(x,y), \lambda (x,y) \st \fname{inA}(x) \qand \fname{inB}(y))\\
\casttoscalar(\alloymath{x}) := (x, \top) \\
\casttoscalar(\alloymath{y}) := (y, \top) \\
\text{then} \\
\casttoscalar(\alloymath{x.h}) := \\
\quad (\lambda y \st h(x,y), \lambda y \st \top \qand \fname{inA}(x) \qand \fname{inB}(y)) \\
\text{and furthermore} \\
\casttoscalar(\alloymath{y.(x.h)}) := \\
\quad (h(x,y), \top \qand \top \qand \fname{inA}(x) \qand \fname{inB}(y))
\end{array}
\end{math}
\end{center} 
If only one operand to a join is a scalar expression, then their join is not a scalar expression. However, an optimization in the next section shows the translation of a join when one operand is a scalar expression.

We can also combine scalars used in an \textbf{if-then-else} expression in the following rule:
\begin{center}
\begin{math}
\begin{array}{rcl}
    \casttoscalar(\alloymath{f => e1 else e2}) &:= & \phantom{mmm} \\
    \multicolumn{3}{r}{(\qite{\trans{\alloymath{f}}}{e_1}{e_2}, \qite{\trans{\alloymath{f}}}{\fname{guard}_1}{\fname{guard}_2})} \\
    \multicolumn{3}{r}{\text{where $\casttoscalar(\alloymath{e1}) = (e_1,\fname{guard}_1)$}} \\
    \multicolumn{3}{r}{\text{and $\casttoscalar(\alloymath{e2}) = (e_2,\fname{guard}_2)$}}
\end{array}
\end{math}
\end{center}
There are also several more ways to combine scalars used in relational expressions shown in Figure~\ref{fig:casttoscalar-relational}.
These rules allow a large class of expressions to be recognized as scalars.

\subsubsection{Translation of Scalar Expressions}

In this section, we describe how a scalar expression is
translated to \fortresslang{} based on its value for the $\casttoscalar$ translation function.
Because $\casttoscalar$ can have a value for a compound scalar
expression (as described in the previous subsection), the
translations in this section are applied at the top-most expression where $\casttoscalar$ has a definition.
At the leaf level, if $\casttoscalar(\alloymath{e}) = (e, \fname{guard})$, then
\begin{align*}
    &\trans{x \in \alloymath{e}} := \fname{guard} \qand x = e \\
    &\trans{(s,x_1,\dots,x_n,r) \in \alloymath{e}} := \\
    &\quad\fname{guard}(s,x_1,\dots,x_n) \qand e(s,x_1,\dots,x_n) = r
\end{align*}
For the Alloy binary operator \alloy{e1 in e2}, where $\casttoscalar(\alloymath{e1}) = (e_1, \fname{guard}_1)$,
\begin{align*}
    &\trans{\alloymath{e1 in e2}} := \qforall x_1: S_1, \dots, x_{n-1}: S_{n-1} \st {} \\
    &\quad\fname{guard}_1(x_1,\dots,x_{n-1}) \qimplies \\
    & \quad\quad \trans{(x_1,\dots,x_{n-1},e_1(x_1,\dots,x_{n-1})) \in \alloymath{e2}}
\end{align*}
This eliminates one level of quantification (the quantification on argument $x_n$ for $\alloymath{e2}$, which is replaced by the value of $\alloymath{e1}$ applied to arguments) compared to the basic translation of \alloy{in} (Section~\ref{sec:formulas}).
Since this translation relies on the translation of a tuple in \alloy{e2}, \alloy{e2} may also be recognized as a scalar expression.

For the Alloy binary operator \alloymath{e1 = e2}, suppose that both operands are scalar expressions with
$\casttoscalar(\alloymath{e1}) = (e_1, \fname{guard}_1)$ and $\casttoscalar(\alloymath{e2}) = (e_2, \fname{guard}_2)$.
Then,
\begin{center}
\begin{math}
\begin{array}{l}
   \trans{\alloymath{e1 = e2}}  :=  \qforall x_1: S_1, \dots, x_{n-1}: S_{n-1}\,\st \\
   \quad (\fname{guard}_1(x_1,\dots,x_{n-1}) \qand \fname{guard}_2(x_1,\dots,x_{n-1}) \\
   \quad \quad {}\qand e_1(x_1,\dots,x_{n-1}) = e_2(x_1,\dots,x_{n-1})) \\
   \quad {}\qor (\qnot \fname{guard}_1(x_1,\dots,x_{n-1}) \qand \qnot \fname{guard}_2(x_1,\dots,x_{n-1}))
\end{array}
\end{math}
\end{center}
If only one operand is a scalar, no special optimization is performed, but the leaf-level optimizations above may be applied within the general translation of \alloy{=}. %

The translation of Alloy join expressions can be optimized if one of the arguments is a scalar.
Suppose that $\casttoscalar(\alloymath{e1}) := (e_1,\fname{guard})$ is a scalar expression of arity $m$ and let \alloy{e2} be any Alloy expression.
Then we can translate:
\begin{align*}
    &\trans{(x_1,\dots,x_n) \in \alloymath{e1.e2}} := \fname{guard}(x_1,\dots,x_m) \\
    &\quad\qand \trans{(e_1(x_1,\dots,x_m),x_{m+1},\dots,x_n) \in \alloymath{e2}}
\end{align*}
Furthermore, if $m=0$ (\myie{} $e_1$ and $\fname{guard}$ take no arguments), then we can translate:
\begin{align*}
    \trans{(x_1,\dots,x_n) \in \alloymath{e2.e1}} := \fname{guard} \qand \trans{(x_1,\dots,x_n,e_1) \in \alloymath{e2}}
\end{align*}
Because this translation relies on the translation of \alloy{e2}, further optimizations can be applied if \alloy{e2} is also a scalar.

\begin{figure*}[h!]
\begin{align*}
    \trans{(x_1,x_2) \in \alloymath{^e}} &:= \fname{guard}(x_1) \qand \Big( x_2 = e(x_1) \qor \Big( \fname{guard}(e(x_1)) \qand \Big( x_2 = e(e(x_1)) \\
    &\qquad\qor\;\Big( \fname{guard}(e(e(x_1))) \qand \Big( \cdots \Big( \fname{guard}(e^{\abs{S}-1}(x_1)) \qand x_2 = e^{\abs{S}}(x_1) \Big) \Big) \Big) \Big) \Big) \Big) \\
    \trans{(x_1,x_2) \in \alloymath{*e}} &:= x_1 = x_2 \qor \trans{(x_1,x_2) \in \alloymath{^e}}
\end{align*}
\caption{Optimized translations of transitive and reflexive closure, where $\casttoscalar(\alloymath{e}) = (e, \fname{guard})$ with arity 1 and $e(x)$ is of sort $S$.}
\label{fig:tc-opt}
\end{figure*}

The optimized translations of transitive or reflexive closure
of an Alloy expression \alloymath{e} of arity two with $\casttoscalar(\alloymath{e}) = (e, \fname{guard})$ are shown in Figure~\ref{fig:tc-opt}. (In this case, we do not use \fortress{}'s built-in closure operators.)
Since $e$ is a unary function (because it is a scalar expression), we can write out the transitive closure of $e$ at $x_1$ as $\{e(x_1), e(e(x_1)), e(e(e(x_1))), \dots\}$.
However, the sort $S$ is finite, so this trace contains at most $\abs{S}$ distinct values: $\{e(x_1), e(e(x_1)), \dots, e^{\abs{S}}(x_1)\}$.
Therefore, it suffices to check that $x_2$ is one of $e(x_1),e(e(x_1)),\dots,e^{\abs{S}}(x_1)$.
However, for any expression $t$, $e(t)$ corresponds to a meaningful Alloy value if and only if $\fname{guard}(t)$ is true.
Hence if $x_2 = e(x_1)$, we require $\fname{guard}(x_1)$; if $x_2 = e(e(x_1))$, we require $\fname{guard}(x_1) \qand \fname{guard}(e(x_1))$; and in general if $x_2 = e^i(x_1)$, we require $\fname{guard}(x_1) \qand \fname{guard}(e(x_1)) \qand \cdots \qand \fname{guard}(e^{i-1}(x_1))$.
Figure~\ref{fig:tc-opt} shows the resulting formula rearranged so that its size is linear in $\abs{S}$.

\textbf{Overall example.}
As an example of the effectiveness of the $\casttoscalar$ abstraction in reducing relations to functions,
suppose $\casttoscalar(\alloymath{f}) = (\lambda (x, y) \st f(x, y), \lambda (x, y) \st g_f(x, y))$ for \alloy{f} a binary scalar expression, and \alloy{a} and \alloy{b} are scalar expressions with $\casttoscalar(\alloymath{a}) = (a, g_a)$ and $\casttoscalar(\alloymath{b}) = (b, g_b)$.
Then \alloy{some b.(a.f)} is translated as follows:
\begin{align*}
    &\trans{\alloymath{some b.(a.f)}} \\
    &=\exists x:S \st \trans{x \in \alloymath{b.(a.f)}} \\
    &=\exists x:S \st g_b \qand \trans{(b,x) \in \alloymath{a.f}} \\
    &=\exists x:S \st g_b \qand g_a \qand \trans{(a,b,x) \in \alloymath{f}} \\
    &=\exists x:S \st g_b \qand g_a \qand g_f(a,b) \qand f(a,b) = x
\end{align*}
\fortress{} performs a further simplification to remove the existential quantifier because its witness ($f(a,b)$) is provided in the formula
based on techniques from \cite{Lampert2017}.

\subsection{Constants-Based Scope Axioms}
\label{sec:constants-scope-opt}

In the general translation, exact and non-exact scopes are translated using set cardinality, which, in turn, is translated as a sum over 
an enumeration of domain elements in the sort.
This method results in large terms and potentially a large scope size 
for integers to cover the maximum set cardinality.

Our \textbf{constants-based scope axioms} method makes direct assertions  about
domain elements being definitely members or not of a sort using the membership predicates.
Suppose \alloy{A} is an Alloy signature with non-exact scope $c$ and let $\sortfn(\alloymath{A}) = S$, \myie{} the sort policy maps \alloy{A} to the \fortresslang{} sort $S$.
To express that at most $c$ domain elements in $S$ may belong to \alloy{A}, we declare $\abs{S}-c$
\fortresslang{} constants
and assert that they are distinct and none of them belong to \alloy{A} using the membership predicate for \alloy{A}.
Therefore, there must be at most $c$ elements in $S$ that are members of \alloy{A}.\footnote{If $c = \abs{S}$, \myie{} \alloy{A} takes up the entire sort $S$, no constants or axioms are added via other parts of the translation and optimizations.}
If \alloy{A} instead has an \textit{exact} scope $c$, then in addition to the above, \portus{} declares $c$ more constants and asserts they are all distinct and members of \alloy{A} using \alloy{A}'s membership predicate.
Combined, these axioms assert that there are $c$ distinct elements in $S$ that are in \alloy{A} and $\abs{S}-c$ distinct elements in $S$ that are not, so there must be exactly $c$ members of \alloy{A}.
This method works even if these axioms are added for multiple subsets of the same set because the method is completely based on using membership predicates, which are distinct for every signature (subset or not).

\section{Evaluation}

We evaluate \portus\ on sets of Alloy models for correctness and completeness, performance of our optimizations, performance relative to \kodkod, and scalability with respect to \kodkod.
All of our performance evaluations are run on a PC with Intel\textsuperscript \textregistered Xeon\textsuperscript \textregistered CPU E3-1240 v5 @ 3.50~GHz with Ubuntu 16.04 64-bit with up to 64~GB of user memory with process limits of 30~GB heap memory and 1~GB stack.
We use Alloy's default solver, \satj~\cite{Berre2010}, in \kodkod\ unless otherwise noted below, and the \zthree~\cite{DeMoura2008} SMT solver in \portus, unless otherwise noted below.
We use the default Alloy semantics for integers (two's complement bit vectors with overflows) 
unless otherwise noted below.
For performance, each run was iterated three times and the average time of the three runs is used in our analysis.   If the first run timed out, the run was not repeated.\footnote{The timeout values used are noted below.}  We clear all caches of the PC between all runs to ensure our runs are independent of the order of runs.

We begin with a \textbf{Benchmark Set} of \numexpert\ Alloy models  scraped from the web selected as models created by experts in a previous study of Alloy models~\cite{Eid2023} because they are complex models created for industrial projects or for peer-reviewed publications.  One of this article's authors (Day) was involved in that study.  This set does not include any models written by the authors of this article.
All of these models were written for Alloy 5, so we apply
a number of scripted transformations to make them compatible with Alloy 6.
Our GitHub repository \texttt{\url{https://github.com/WatForm/portus-evaluation}} contains scripts that download and transform to Alloy 6 the models of the Benchmark Set. This repository also contains our Feature Model Set (described in Section~\ref{sec:scalability} below), 
and it contains scripts to reproduce our evaluation results.

\subsection{Completeness and Correctness}

\resques{How complete is our translation?}
We evaluate how many queries are supported by \portus\ and \kodkod\ for all \numexpert\ models and all commands in the Benchmark Set for a total of \numqueriestotal\ queries.
Our implementation of \portus\ is able to execute all but \numqueriesportuscantdo{} of the \numqueriestotal\ queries.
\Numberstringnum{\numqueriesportuscantdosecondorder} of the queries use second-order quantifiers in unsupported ways as described in Section~\ref{sec:unimplemented}.
\Numberstringnum{\numqueriesportuscantdounivfield} others 
contain a field with the bounding expression \mbox{\alloy{univ}} and \numberstringnum{\numqueriesportuscantdobadjoin} contain a join with both integers and other atoms in the joined column; these were the only cases in the Benchmark Set where a problem was caused by an expression not having definite sorts.
The \alloyanalyzer\ is able to execute all but \numberstringnum{\numquerieskodkodcantdo} query in our Benchmark Set, which is a query whose model includes a second-order quantifier that the \alloyanalyzer{} is unable to skolemize but which \portus{} supports.\footnote{We have filed a bug report for the \alloyanalyzer{} at \texttt{\url{https://github.com/AlloyTools/org.alloytools.alloy/issues/326}}.}
Excluding the models with queries that are not supported by both methods, we are left with \numsupport\ models and 337 queries.

We also attempt to run existing implementations of related work (described in Section~\ref{sec:related}) on translations from Alloy to SMT solvers.  The only existing, available implementation that we could find is 
\alloytosmt~\cite{Mohamed2019}, which translates Alloy to the smt2 format, but depends on the \cvcfour~\cite{Barrett2011} native input language~\cite{CVC4Native}. On our Benchmark Set of \numexpert\ models, \alloytosmt\ was able to translate 37 of them to an output smt2 file.  Of these files, 33 ran without an error in \cvcfour\ with a total of 86 queries (check-sats in SMT).  Of these queries, 11 are satisfiable, 11 are unsatisfiable, and 64 reported unknown in \cvcfour. 
We describe the differences between our work and all previous translations from Alloy to SMT solvers in Section~\ref{sec:related} (related work).
From our empirical evaluation (described above) and our analysis of the literature, we claim that we currently have the most complete working implementation of a translation from Alloy to an SMT solver.

\resques{Is our implementation of \portus\ producing correct results?} %
In addition to extensive unit testing to check the correctness of \portus, we use Jazzer~\cite{Jazzer} fuzz testing repeatedly after most major changes to the code to generate 
well-formed Alloy abstract syntax trees
and we fixed a number of bugs that were found by this technique.  Jazzer can generate a different set of tests in every iteration. 

Next, we evaluate the correctness of the \totalcorrectnessqueries\ queries in the Benchmark Set supported by both \portus\ and \kodkod.
For each query, we run both \portus\ and the \alloyanalyzer\ with the \kodkod\ solver on the model and check that they both return the same satisfiability result (SAT or UNSAT).
There are \totalcorrectnessqueriessat\ satisfiable queries and \totalcorrectnessqueriesunsat\ unsatisfiable queries in the Benchmark Set according to the \alloyanalyzer.

For satisfiable queries, we additionally check that the solution returned by \portus\ is correct.
Since there are, in general, many solutions to a satisfiable query, the solution returned by \portus\ may not be identical to the first solution returned by \kodkod, so we use the following process to ensure \portus's solution is valid according to \kodkod.
We transform \portus's solution into an Alloy \texttt{A4Solution} object representing the same interpretation of the signatures and fields in the model.
Using the \alloyanalyzer's built-in translator from Alloy to \kodkod, we then extract the \kodkod\ formula that is sent to the \kodkod\ solver when running the \alloyanalyzer\ on the query.
This is the formula used by the \kodkod\ solver to generate its satisfiability result for the query; it is generated completely independently of \portus\ and contains all information relevant to the query, including the command, facts, signature hierarchy, and field constraints.
The \alloyanalyzer\ contains a built-in evaluator which can evaluate whether a given \kodkod\ formula holds in a given \texttt{A4Solution}. Using the evaluator, we check that the \kodkod\ formula holds in the \texttt{A4Solution} representing \portus's solution.\footnote{This evaluation time is not included in our performance evaluations because it is only necessary for correctness checking.}
This correctness testing process is performed programmatically and  ensures that the solution returned by \portus\ is a valid instance of the query according to \kodkod.

There are \numqueriesportustimeout\ queries in four %
models which \portus{} does not solve within a timeout of 72 hours with 70 GB of user memory. %
On the remaining \totalcorrectnessqueriesfinished\ queries that \portus\ solves, we find that \portus\ correctly gives the same SAT/UNSAT determination as \kodkod{}
(\numcorrectnesssat\ are satisfiable and \numcorrectnessunsat\ are unsatisfiable).
The \alloyanalyzer's evaluator does not support second-order quantifiers, so we could not perform the solution correctness check for the \numberstringnum{\numquerieswithsecondorder}\ supported Alloy queries containing second-order quantifiers, but for the remaining 180 satisfiable queries, \portus's solution is verified to be correct using the method described above.

There is a slight discrepancy between \kodkod's and \fortress{}'s semantics for finite integers. \fortress{} uses a technique called overflow-preventing finite integers (OPFI)~\cite{Zila2023}.
In some cases, OPFI does not produce an overflow when \kodkod{} does.
Additionally, \fortress{} uses the underlying SMT solver's integer semantics which may not match \kodkod's semantics for divide-by-zero and mod-by-zero.
For consistent results,
we modify \portus{} to mimic \kodkod's integer results during correctness testing only.\footnote{
For one model (the Chord model \cite{Zave2012a}), we use the ``Prevent overflows'' option in \kodkod{} for integers, and we increase the bitwidth for some commands, because this option is necessary in Chord for \kodkod{} to give the known correct result for all commands as specified by \cite{Zave2012a} and the model itself.
}

\subsection{Performance of Portus Optimizations}

\resques{Do the \portus\ optimizations boost its performance?}
Each model in the Benchmark Set may contain multiple (uncommented out) run/check commands. 
Randomly, we choose one  
command for each of the \numsupport{} models in the Benchmark Set that is supported by both \portus{} and \kodkod{} to include in our \textbf{Performance Evaluation Set (PE Set)}.
Our PE Set thus consists of \numsupport{} queries, each in a different model.
We choose only one command per model to avoid having a model with many queries bias the aggregate performance results.
There are \numsat{} satisfiable and \numunsat{} unsatisfiable queries in this set.

\begin{figure}[h]
\includegraphics[width=\columnwidth]{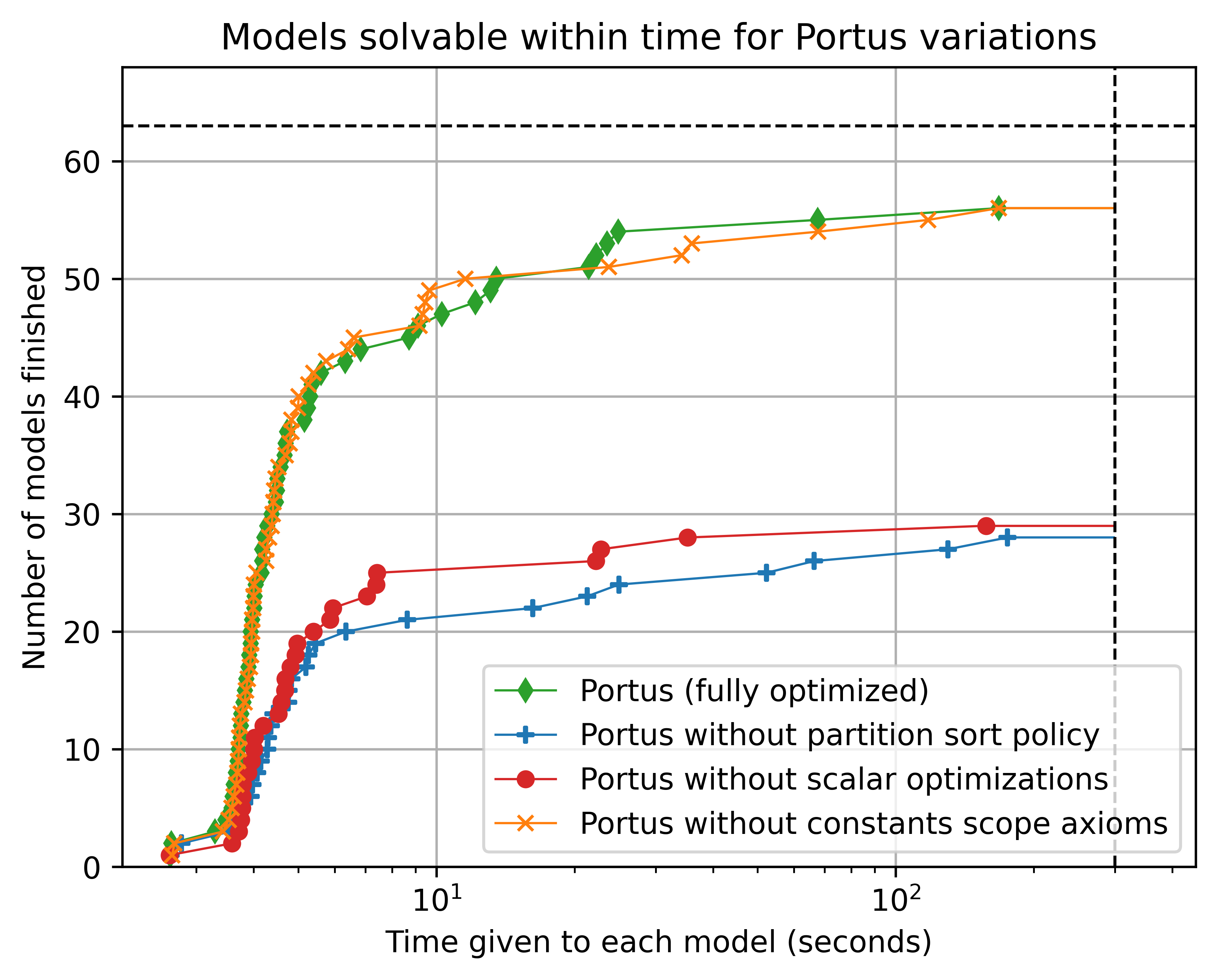}
\caption{Performance of \portus\ optimizations}
\label{fig:eval-opt}
\end{figure}

To evaluate the optimizations presented in Sections~\ref{sec:partition-sort-policy}--\ref{sec:constants-scope-opt}, we run the queries in our PE Set in four configurations of \portus: 
\begin{enumerate}
\item fully optimized %
\item all optimizations except the partition sort policy
\item all optimizations except the scalar optimizations 
\item all optimizations except the constants-based scope axioms
\end{enumerate}
Figure~\ref{fig:eval-opt} shows a graph of how many queries are solved within a timeout of five minutes.
The fully optimized version of \portus\ solves 56 (of \numsupport) queries within five minutes. From this graph, we conclude that the fully optimized version of \portus\ completes more queries in less time than the other methods.  
The version of \portus{} using cardinality-based rather than constants-based scope axioms completes the same number of queries (56) as the fully optimized version, and it completes the 56th model in approximately the same amount of time, but it completes the 55th model in 117.7 seconds while the fully optimized version takes only 67.6 seconds.

\subsection{Performance compared to \kodkod}

\begin{figure}[h]
\includegraphics[width=\columnwidth]{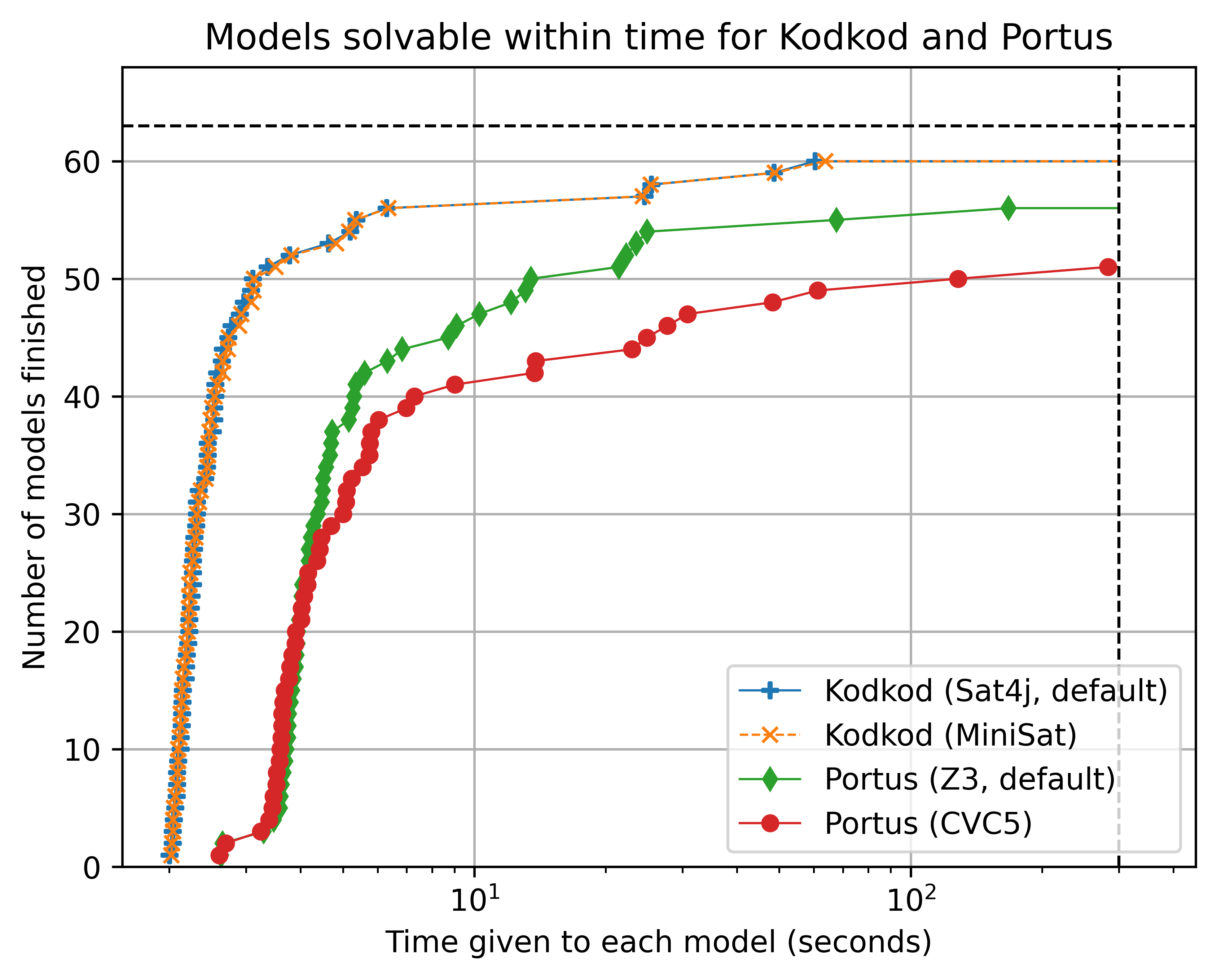}
\caption{Performance of \portus\ vs \kodkod}
\label{fig:eval-kodkod-portus}
\end{figure}

\resques{How does \portus\ compare to \kodkod\ in performance?}
Using the \numsupport{} queries of the PE Set,
we evaluate the fully optimized version of \portus\ with both the \zthree~\cite{DeMoura2008} solver and the \cvcfive~\cite{Barbosa2022} solver in comparison to the \alloyanalyzer\ with the \kodkod\ solver combined with both \satj~\cite{Berre2010} and \minisat~\cite{Een2004}.  
Figure~\ref{fig:eval-kodkod-portus} shows a graph of how many queries are solved by each solving method within a timeout of five minutes (dashed vertical line). 
\kodkod\ outperforms \portus\ but not by a large amount.  
For example, \kodkod{} with \satj\ solves 56 queries and \portus{} with \zthree\ solves 50 within twenty seconds. 
There are three queries not solved by \kodkod{} and seven queries not solved by \portus{} within the timeout.
For one query, \texttt{serializableSnapshotIsolation.als}, 
\portus{} completes in 22.3 seconds whereas \kodkod{} takes 911.7 seconds, which provides motivation for having a portfolio of solvers available in the \alloyanalyzer.

From Figure~\ref{fig:eval-kodkod-portus}, it also can be seen that \kodkod{} with \satj\ and \minisat{} perform 
equivalently
and that \portus{} with \zthree\ performs better than with \cvcfive, justifying the choice of \satj{} and \zthree{} as the defaults.
 In other contexts~\cite{AlloyQuickGuide2025}, \minisat{} may have better performance than \satj.

\subsection{Scalability}
\label{sec:scalability}

\newcommand*{\scalingtimeout}{five} %

\resques{Are there models in which  \portus\ is more scalable over scope sizes compared to \kodkod?}
We 
create a set of problems that are \textit{hard} for \kodkod{} with respect to scaling 
the scope of some signature in the model.
For each query in the PE Set, we run \kodkod\ once on a series of queries
that each use an exact scope size in the set $\{2,4,6,8,10, 12, \dots, 80\}$
for every top-level, non-one, non-lone, non-enum signature.  When scaling a signature, we
keep the other signatures of the model at their values given in the command.
Exact scopes are used to force the instances to
be larger during scaling.  
We record the maximum (up to 80) scope size for each signature 
for which \kodkod{} completes within a \scalingtimeout\ minute timeout
from this evaluation.\footnote{At this point, we had to remove one query from the PE Set because the universe grows too large for \kodkod{} (but not \portus) to represent a high-arity field before timing out.}
We then determine which signature of the model has the \textit{lowest} of these maximum
scope sizes.\footnote{If there is a tie between the scope size of two signatures, we randomly choose one to scale for our results.}
The chosen signature
scales the \textit{worst} for the query
according to \kodkod. %
During scaling, we use the scope values found in the command
(or defaults) for all other signatures of the model.
By scaling only one signature in the model, 
our goal is to emphasize the scaling of the set that makes the problem hard 
for \kodkod{}
and to limit the 
bias due to models having more signatures.

\begin{figure}[t]
\includegraphics[width=\columnwidth]{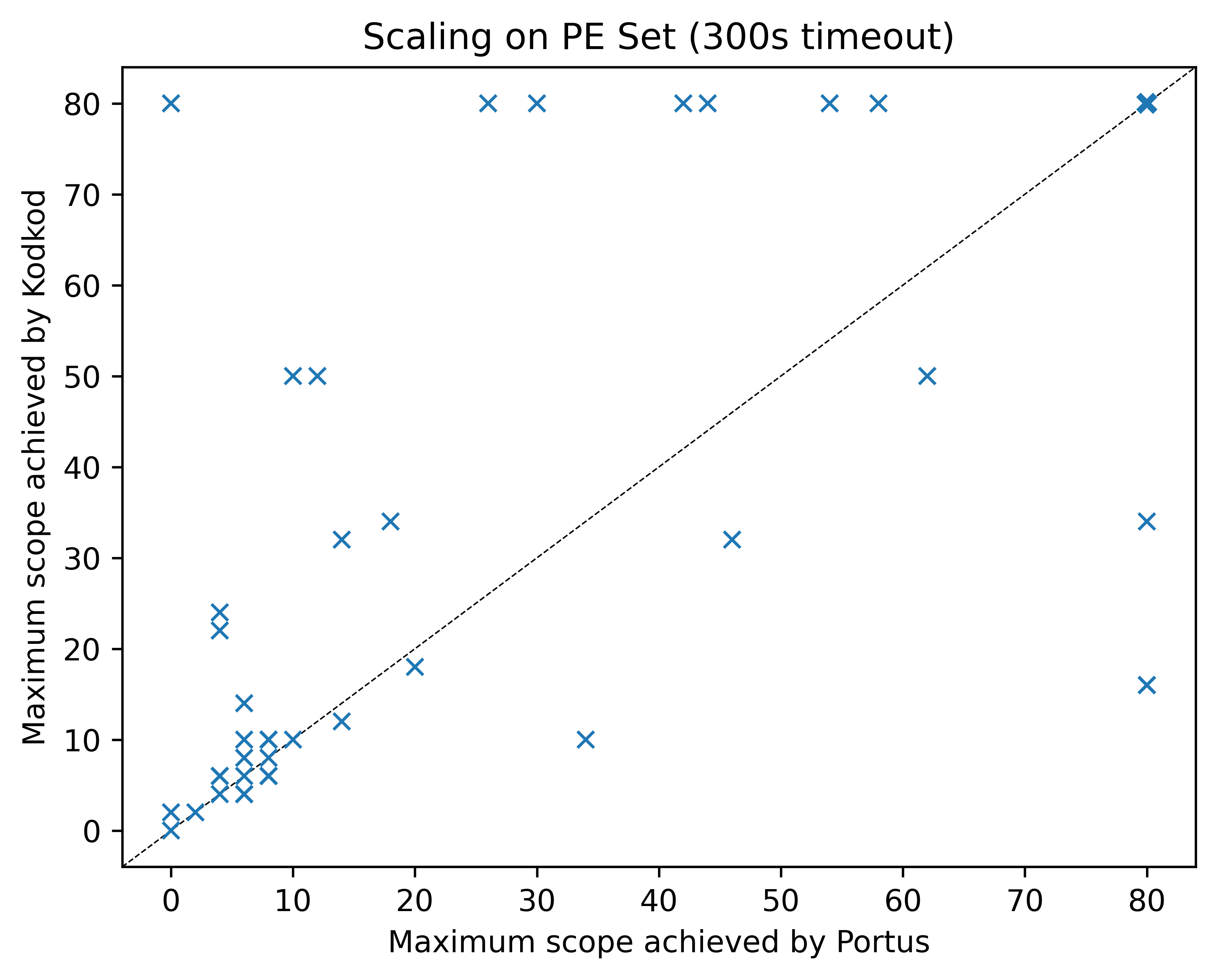}
\caption{Scalability of \portus\ vs \kodkod{} (some points overlap)}
\label{fig:eval-kodkod-portus-scalability}
\end{figure}

Next, for each model in our PE Set, 
we run \portus{} once on a series of problems that use an exact scope size in the set $\{2,4,6,8,10,12,\dots,80\}$
for the chosen worst-scaling signature only.
We again use a timeout of \scalingtimeout{} minutes %
and we record the maximum scope size for which \portus{} completes within this timeout.
If \portus{} achieves a higher maximum scope size within the timeout than \kodkod{} does for a particular model, then
\portus{} scales better than \kodkod{} on the chosen \textit{hard} signature.
Figure~\ref{fig:eval-kodkod-portus-scalability} plots a point for each query at $(x,y)$ where $x$ is the maximum scope size achieved by \portus{} and $y$ is the maximum scope size achieved by \kodkod{}.  
Points above the dashed line are queries where \kodkod{} scaled better than \portus{} and vice versa. 
In this figure, points overlap; in particular the upper right `$\times$' represents 24 queries.
Of the 39 queries that reach a scope of 80 before the timeout, there are twelve queries where \portus{} scales better than
\kodkod.

\resques{Is there evidence that \portus\ might scale better on some Alloy language features than \kodkod?}
It would be useful to know if \portus\ is likely to perform better than \kodkod\ on models that have specific language features.  In this section, we describe a preliminary investigation toward answering the question of when to use \portus\ and when to use \kodkod.  We manually construct some very simple Alloy models that try to isolate an Alloy language feature (\myeg\ transitive closure) and that show differences in the scalability of the model in \portus\ and \kodkod.
We call this set of models our \textbf{Feature Model Set (FM Set)}. 
These models are all satisfiable.  
We scale each signature of each model with scope values in the set $\{2,4,6,8,10, 12, \dots, 80\}$ with a timeout of five minutes.
For models with multiple signatures, the other signatures are kept at a fixed scope. After a timeout occurs, we stop increasing the scope.
The results of this evaluation for some models in our FM Set are shown in Figure~\ref{fig:eval-kodkod-portus-feature-model-scalability} and the simple models are shown in the captions of 
the figure.

\newcommand*{\scalefigwidth}{0.31\textwidth}

\afterpage{
\onecolumn
\begin{figure*}
\begin{center}
\subfloat[Functions, scaling the domain (\alloy{A}, left) and the range (\alloy{B}, right). The scope of the signature not being scaled is fixed at 16. \\ \alloy{sig A \{ f: A->A->one B \}; sig B \{\}; run \{\}}]{
    \label{fig:scale-functions}
    \hspace*{1.8cm}
    \includegraphics[width=\scalefigwidth]{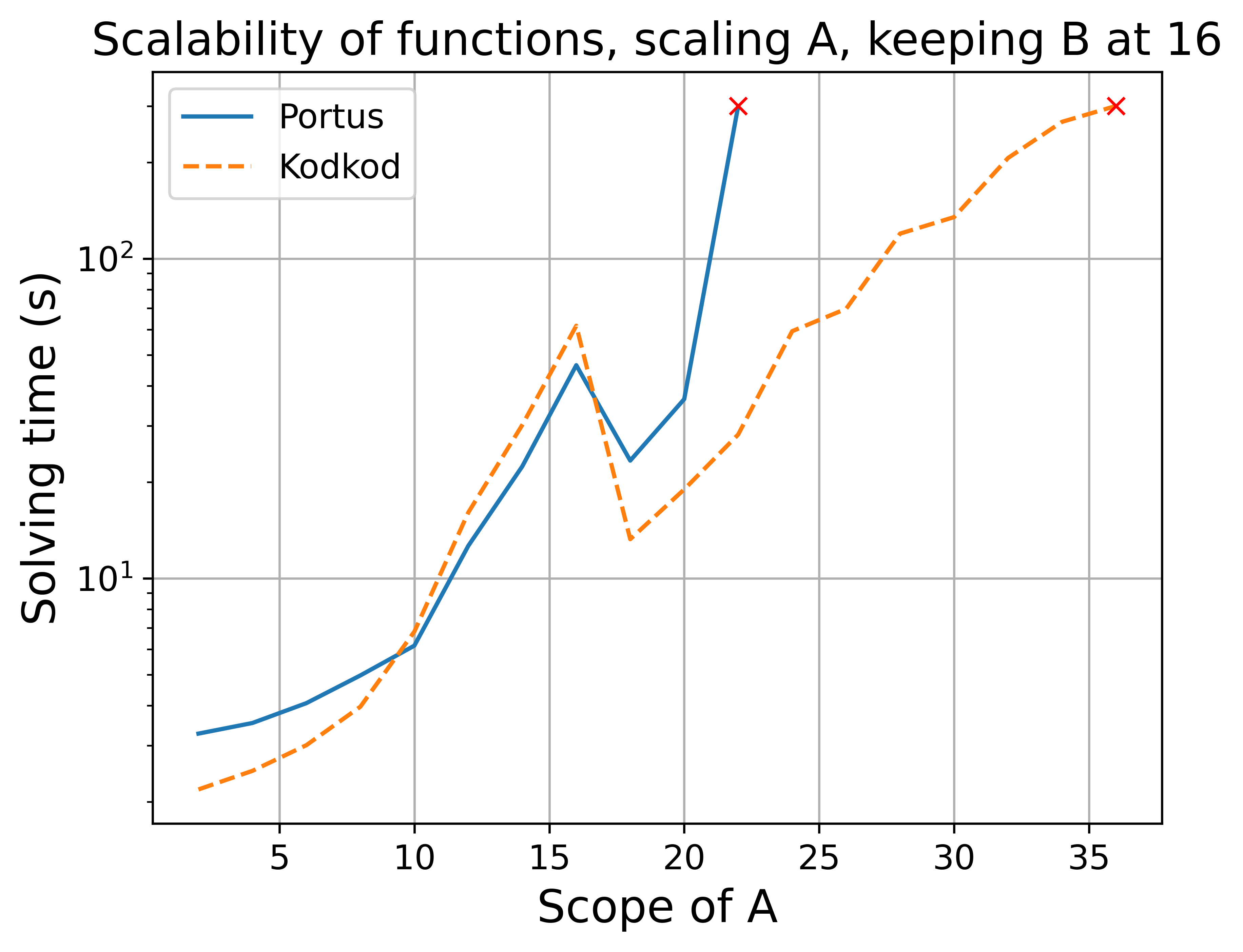}
    \hfill
    \includegraphics[width=\scalefigwidth]{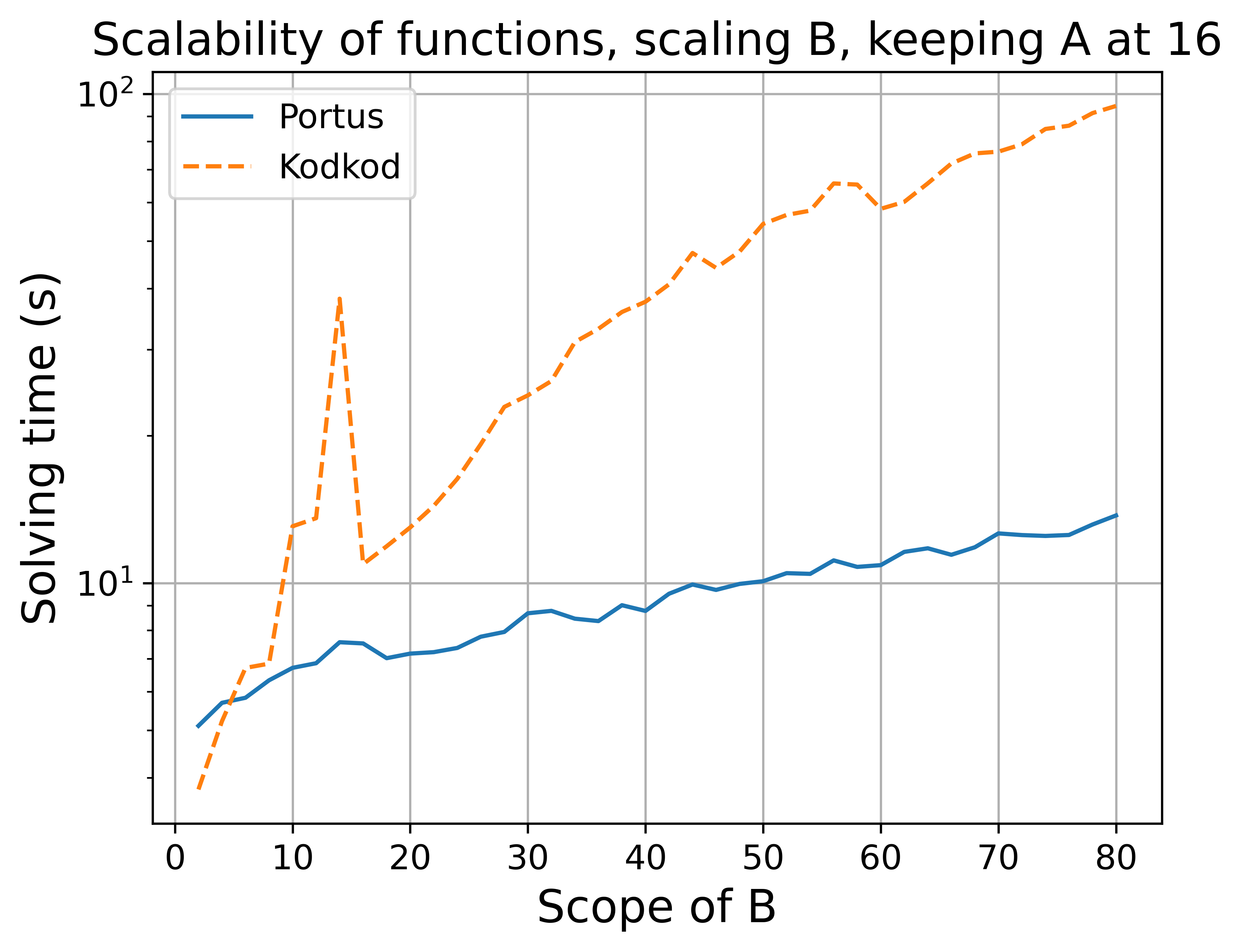}
    \hspace*{1.8cm}
}
\hfill
\subfloat[Functions at higher scopes, scaling the domain (\alloy{A}, left) and the range (\alloy{B}, right). The scope of the signature not being scaled is fixed at 32. \\ \alloy{sig A \{ f: A->A->one B \}; sig B \{\}; run \{\}}]{
    \label{fig:scale-functions-higher-scope}
    \includegraphics[width=\scalefigwidth]{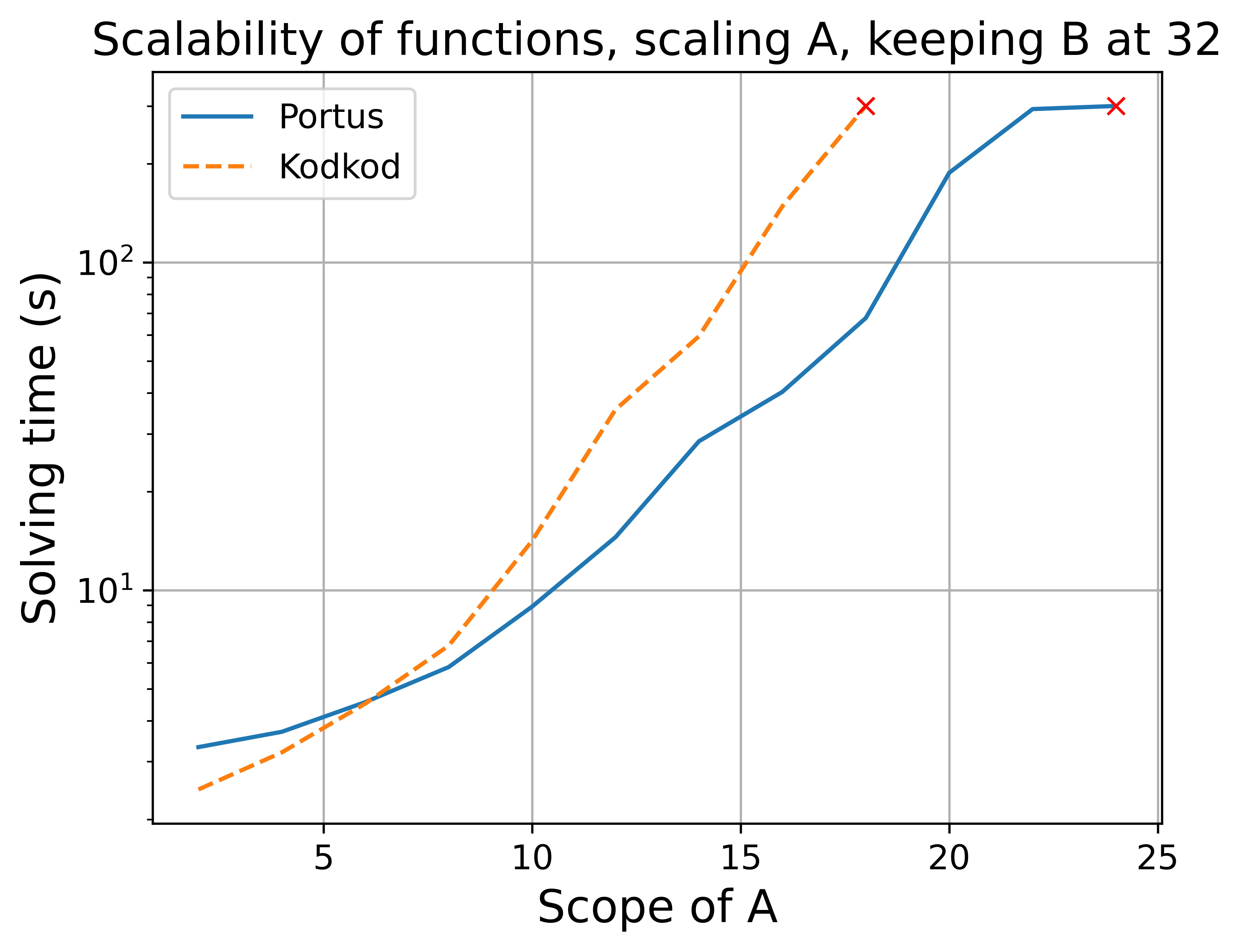}
    \hfill
    \includegraphics[width=\scalefigwidth]{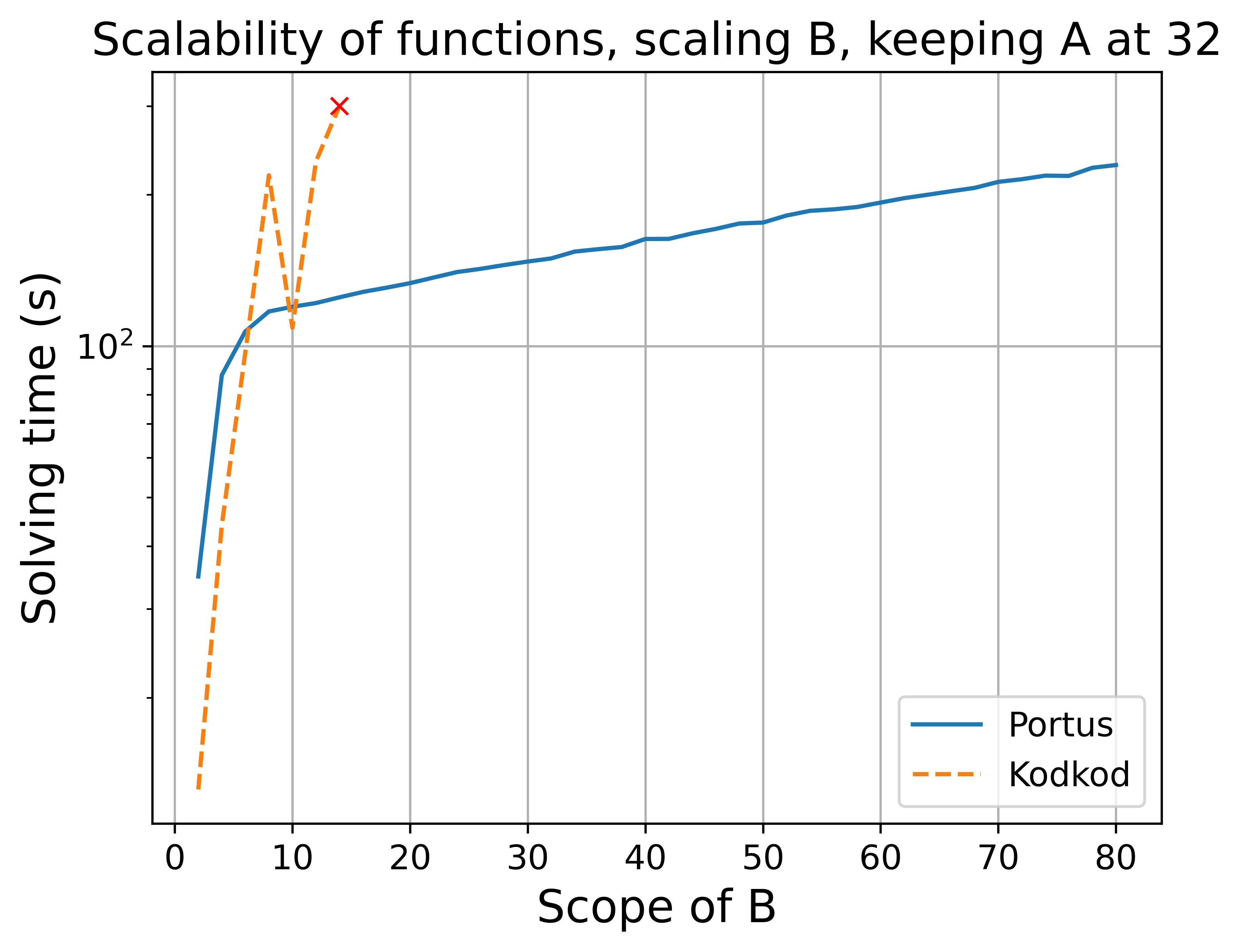}
}
\hfill
\subfloat[Function composition, scaling \alloy{A}. \\ \alloy{sig A \{ f: A, g: A \}} \\ \alloy{run \{ some f.g \}}]{
    \label{fig:scale-function-composition}
    \includegraphics[width=\scalefigwidth]{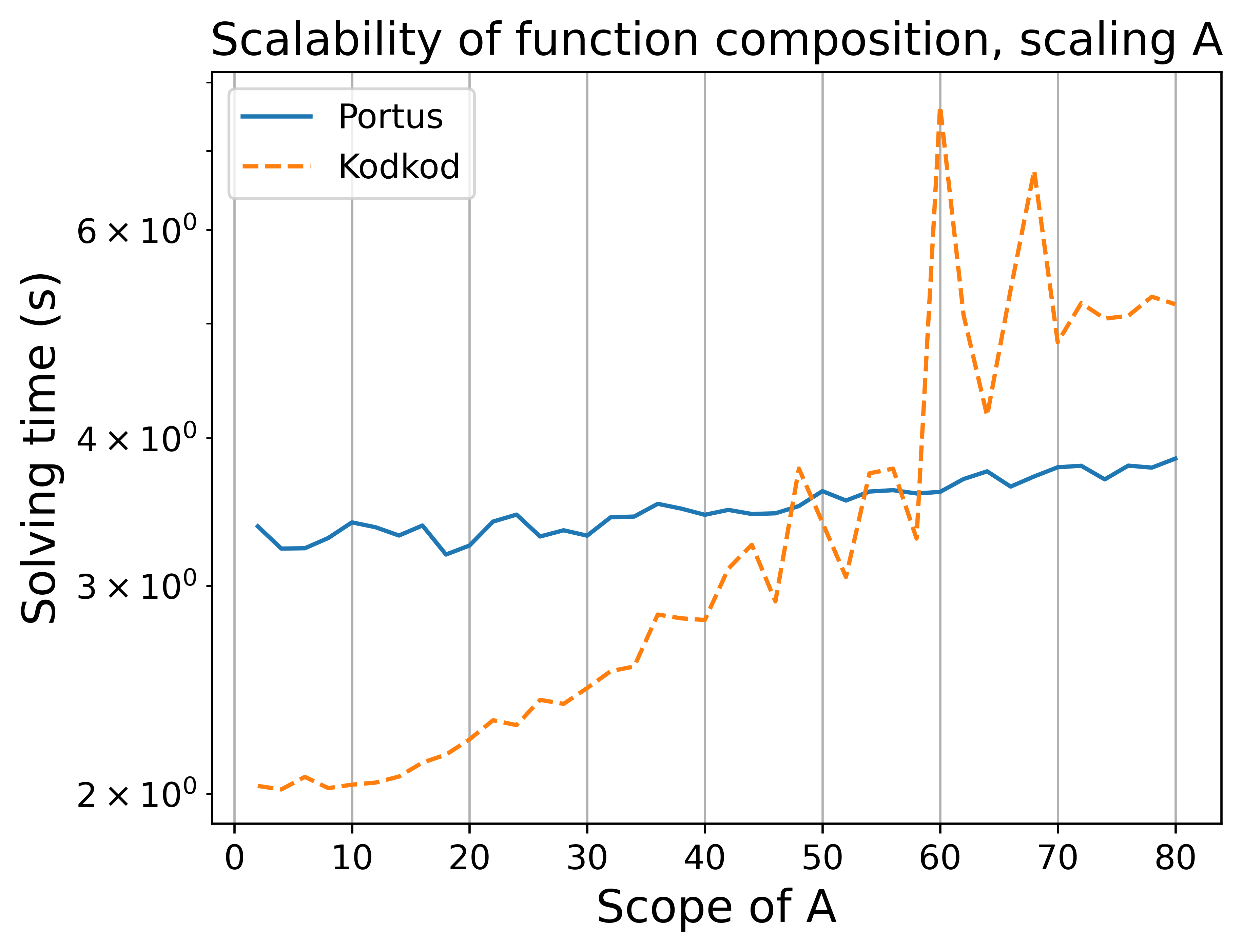}
}
\hfill
\subfloat[Relations, scaling the domain (\alloy{A}, left) and the range (\alloy{B}, right). The scope of the signature not being scaled is fixed at 16. \\ \alloy{sig A \{ f: A->A->B \}; sig B \{\}; run \{\}}]{
    \label{fig:scale-relations}
    \includegraphics[width=\scalefigwidth]{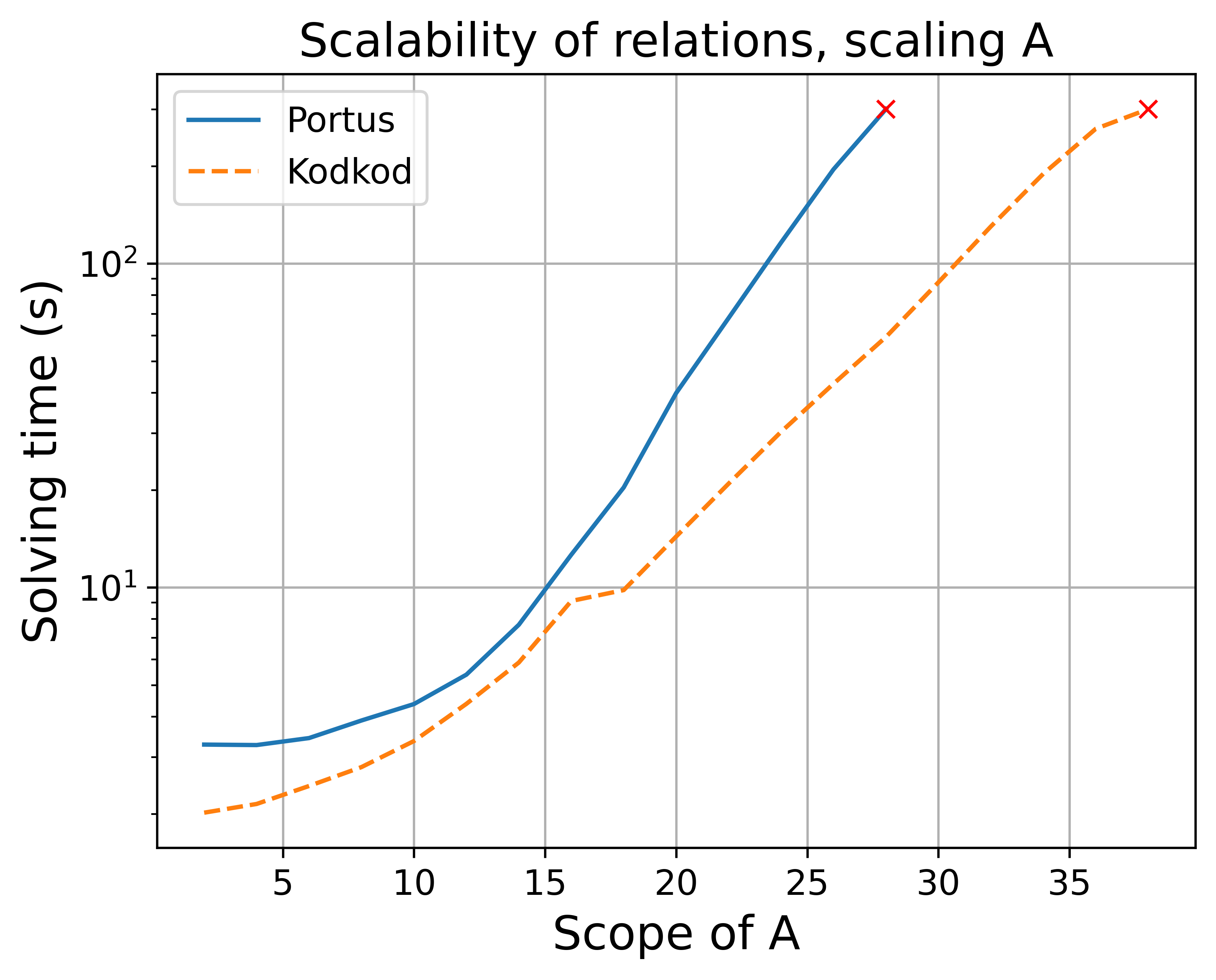}
    \hfill
    \includegraphics[width=\scalefigwidth]{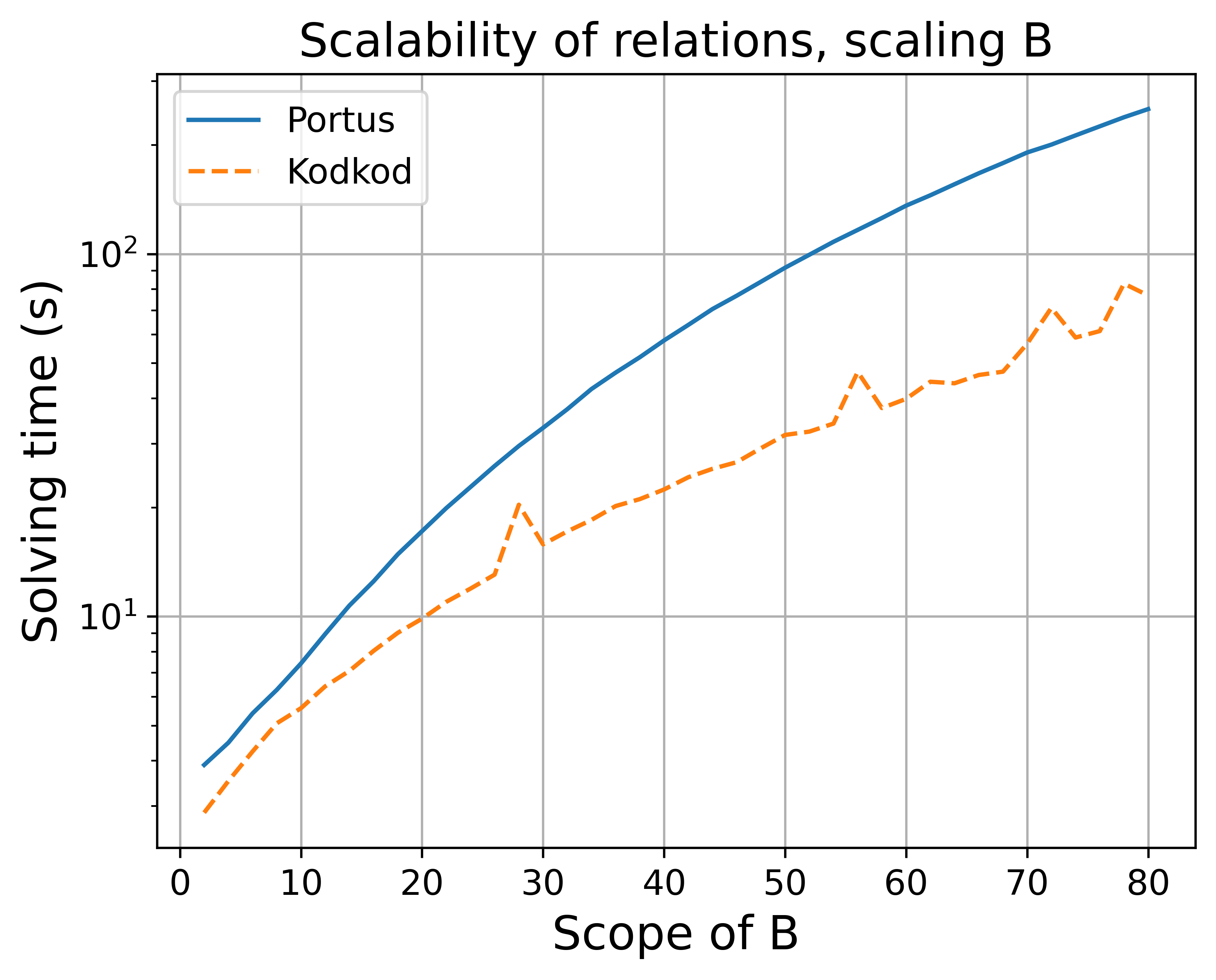}
}
\hfill
\subfloat[Ordering module, scaling \alloy{A}. \\ \alloy{open util/ordering[A]; sig A \{\}} \\ \alloy{run \{ all a: A | lte[first, a] \}}]{
    \label{fig:scale-ordering}
    \includegraphics[width=\scalefigwidth]{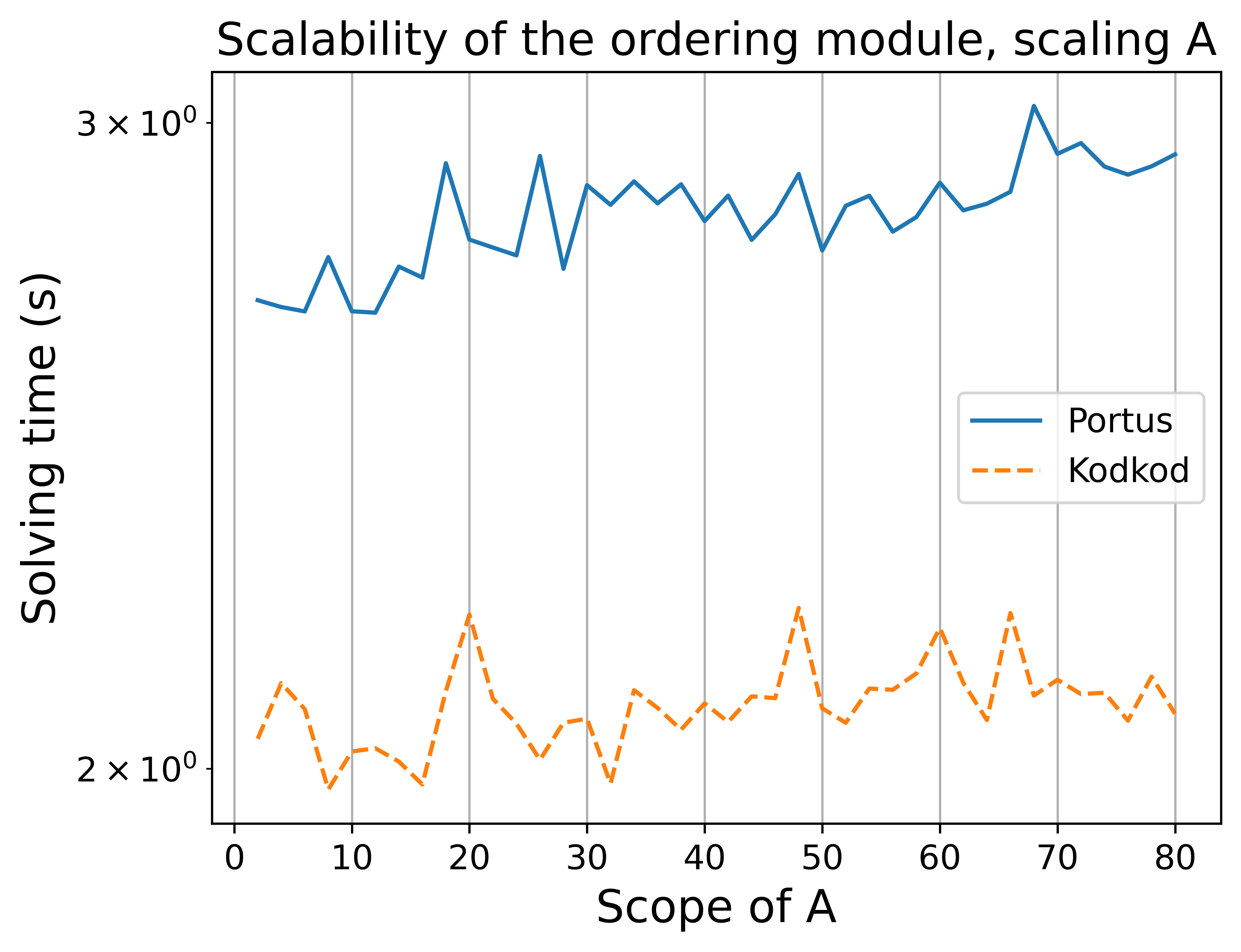}
}
\hfill
\subfloat[Transitive closure over a function, scaling \alloy{A}. \\ \alloy{sig A \{ f: A \}} \\ \alloy{run \{ some ^f \}}]{
    \label{fig:scale-tc-function}
    \includegraphics[width=\scalefigwidth]{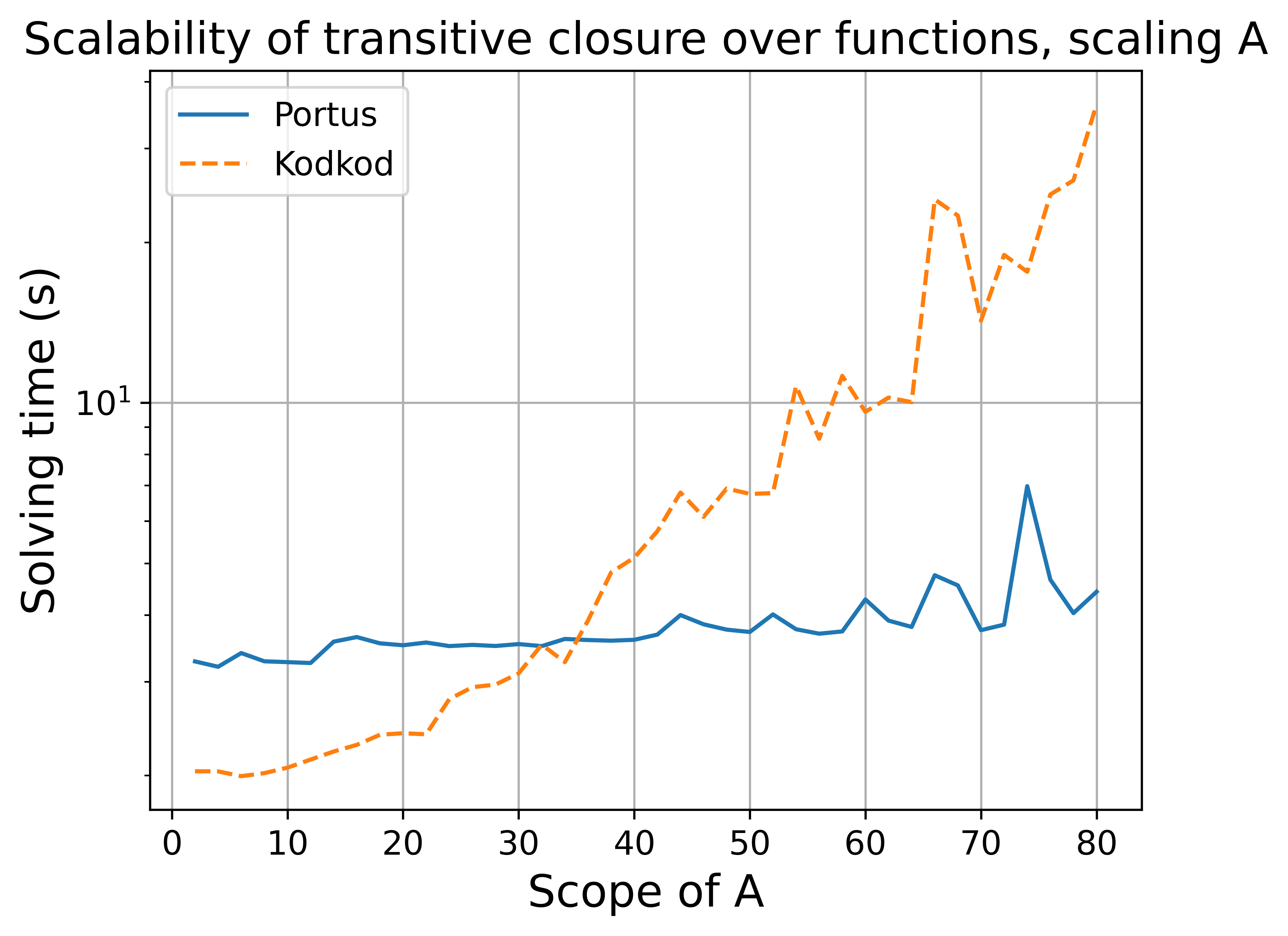}
}
\hfill
\subfloat[Transitive closure over a relation, scaling \alloy{A}. \\ \alloy{sig A \{ f: set A \}} \\ \alloy{run \{ some ^f \}}]{
    \label{fig:scale-tc-relation}
    \includegraphics[width=\scalefigwidth]{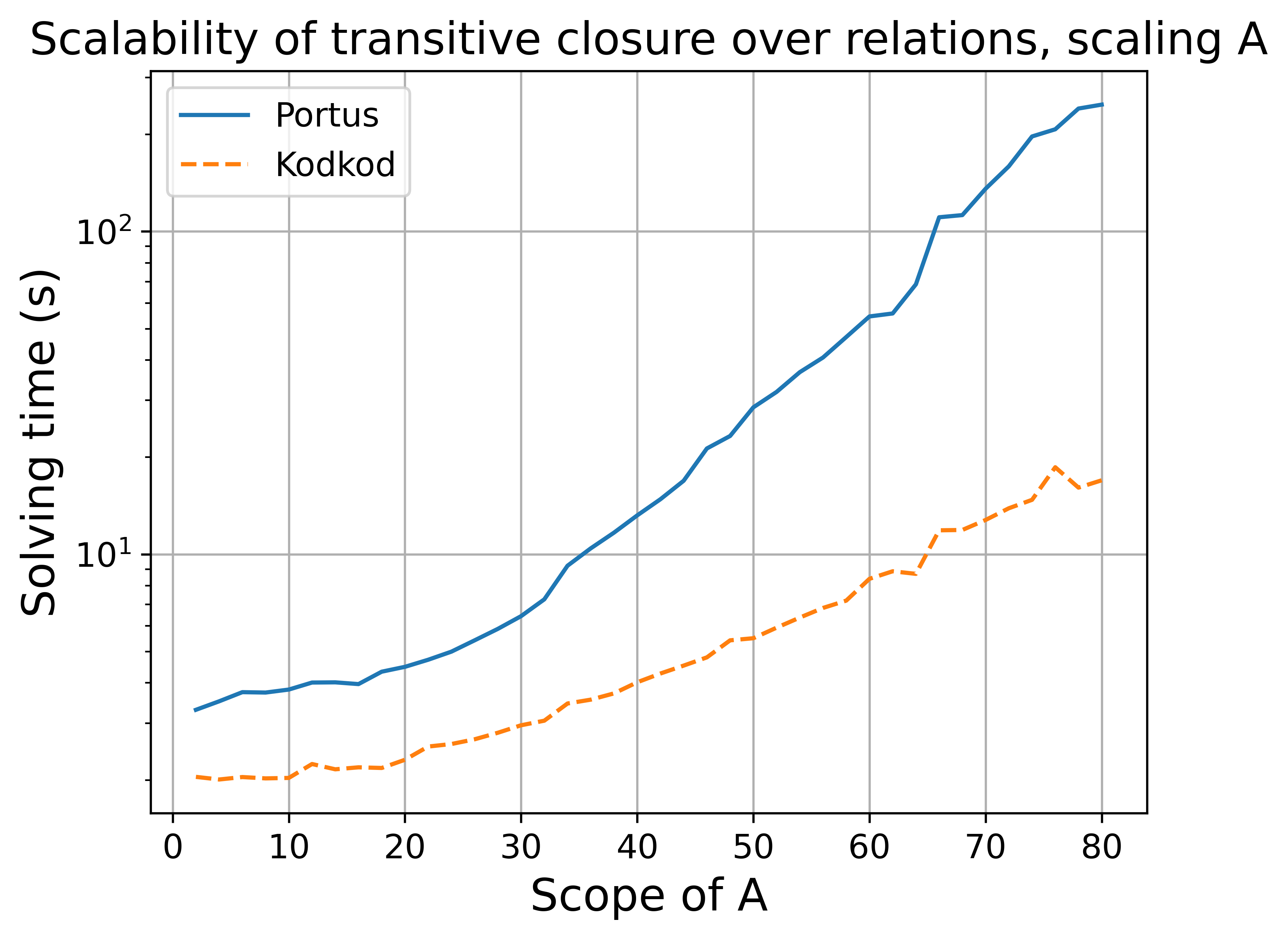}
}
\hfill
\subfloat[Cardinality, scaling \alloy{A}. \\ \alloy{sig A \{\}} \\ \alloy{run \{ \#A > 0 \} for 10 Int}]{
    \label{fig:scale-card}
    \includegraphics[width=\scalefigwidth]{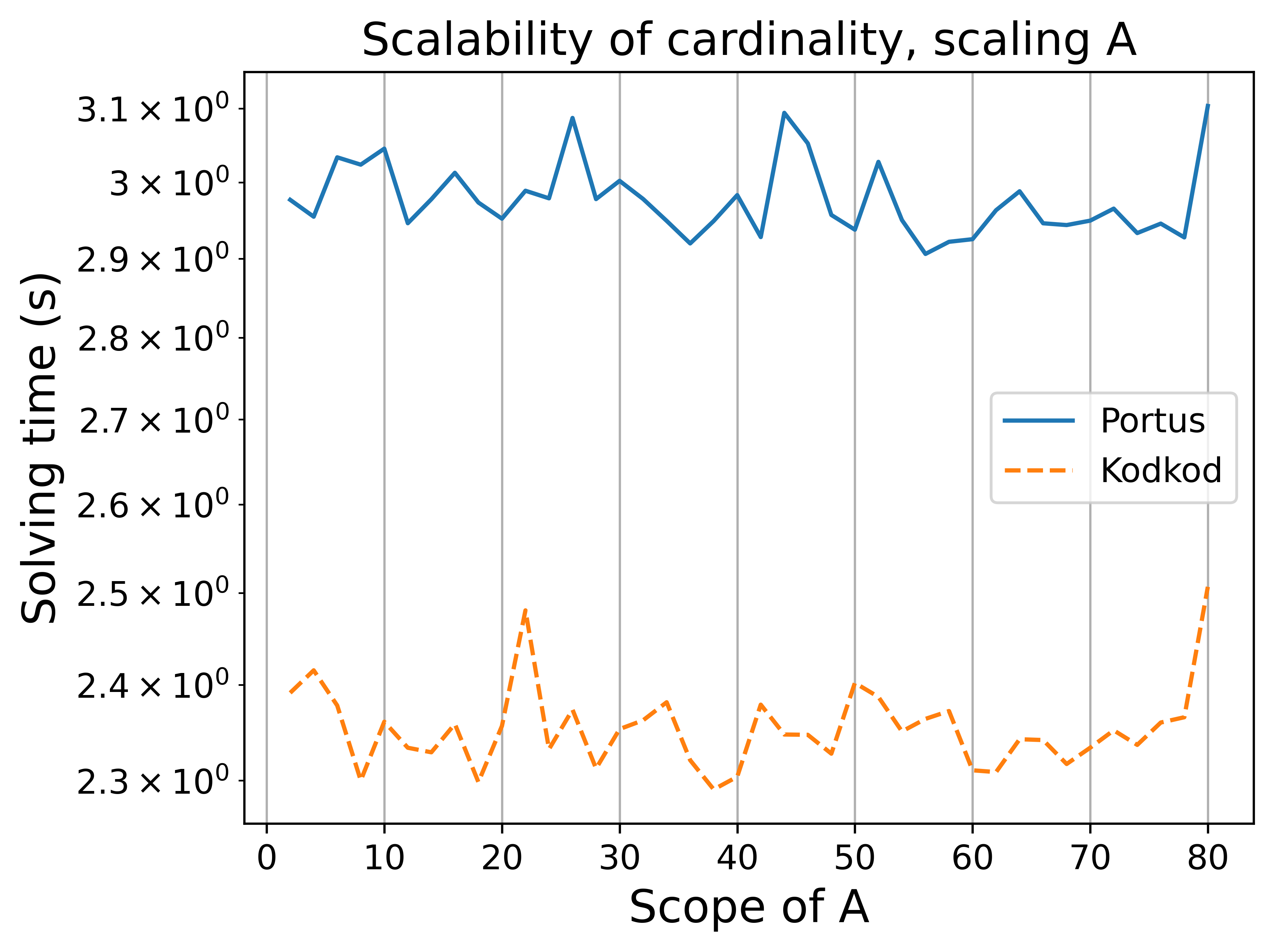}
    }
\end{center}
\caption{Scalability of \portus\ vs \kodkod\ on specific features.\protect\footnotemark
A red `$\times$' denotes a timeout at five minutes. }
\label{fig:eval-kodkod-portus-feature-model-scalability}
\vspace*{-2cm}
\end{figure*}
\footnotetext{Functions and relations of high arity, \myeg{} \alloy{f: A->A->one B}, were used in Figs.\ \ref{fig:scale-functions}, \ref{fig:scale-functions-higher-scope}, and \ref{fig:scale-relations} so that the trend in how \portus{} and \kodkod{} scale relative to each other is clearly visible within the prescribed scope size.
}
\twocolumn
}

First, Figure~\ref{fig:scale-functions} shows that for a simple Alloy model with a field that is a function,
scaling the signature in the range of such an Alloy function produces a particular advantage for \portus, likely because \fortress{} can apply more symmetry breaking constraints in this case~\cite{Poremba2023}.
Figure~\ref{fig:scale-functions-higher-scope} shows this effect is especially pronounced when the domain is fixed at a higher scope.
Figures~\ref{fig:scale-function-composition} and \ref{fig:scale-tc-function} are meant to exemplify composition of functions and transitive closure over a function, respectively, and show that these models scale extremely well in \portus{} in comparison to \kodkod.
Figures~\ref{fig:scale-relations} and \ref{fig:scale-tc-relation} show that non-function (relation) Alloy fields and transitive closure in these models, respectively, scale less well in \portus.
We conjecture that when the scalar optimizations from Section~\ref{sec:scalar-opt} can be applied, \portus{} scales well.
We also tried scaling models that include the ordering module and set cardinality (Figures~\ref{fig:scale-ordering},~\ref{fig:scale-card}) and our results show that \kodkod{} scales much better for these models.

While this language feature evaluation is not comprehensive, it does provide evidence that some language features might have better performance with respect to scalability in \portus\ compared to \kodkod.
Our analysis provides hints of where future evaluation can explore a systematic way to determine which method (\portus\ or \kodkod) would be better for certain language features found in a model.

\subsection{Threats to Validity.}

There are a number of threats to the validity of our correctness and performance results. 
\textbf{External Validity.} 
Our Benchmark Set consists of models made available on the web and judged manually in a previous work~\cite{Eid2023} (where Day was also an author) to be Alloy models written by experts.  
The commands randomly chosen from these models were all uncommented commands in the file.  
In some cases, there are comments in the model text regarding commented out Alloy commands that the authors of the model found to have performance issues for Alloy. 
This set of models may not be representative of all Alloy models and thus our performance results may differ on a different set of Alloy models. 
However, as these models are available publicly on the web, we believe it is fair to assess the uncommented-out commands in these models as likely to complete in Alloy within a time that the user is willing to wait for results, thus potentially making our Benchmark Set favourable to \kodkod{} for performance, whereas other models or alternative ways to model behaviour (such as using functions rather than a relation with an extra constraint) might give an advantage to \portus{} in performance.
Our Feature Model Set of toy models was created manually and does not cover all language features in Alloy. Each model in the set was created to exemplify one language feature; however, many language features cannot be used in isolation, so some models use additional language features (\myeg{} quantifying expressions with \alloy{some}) to produce a model using the desired feature. The results in Figure~\ref{fig:eval-kodkod-portus-feature-model-scalability} therefore only suggest how the language features may scale, as the additional language features used in the models may influence the scaling result.
Other models could show evidence of language features where \kodkod\ scales better than \portus.
\textbf{Internal Validity.} 
Our correctness evaluation of \portus{} is relative to \kodkod{}'s results.  
Our method does recognize that there may be more than one satisfiable solution and thus does not look for \portus{} and \kodkod{} to produce the same solution.  
We did not evaluate the correctness of the scalability results.  
We did extensive unit and fuzz testing to ensure the correctness of our implementation, but more testing is always beneficial.
Our performance evaluation was careful to measure solving time with cleared caches and as the only user-application process running on the server.
Alternative SAT solvers with \kodkod{} or SMT solvers with \portus{} may produce different relative performance results.
\textbf{Construct Validity.} 
Our comparison of the correctness and performance of \portus{} to \kodkod{} looks at queries in general for 
the Benchmark Set and then uses a constructed Feature Model Set to isolate language features.
More detailed research questions could look at
differences in performance based on the satisfiability status of models, scope of particular sets, and more language features. 
The choice of scope can affect the satisfiability of the model, which can affect scalability.

\section{Related Work}

\label{sec:related}

\begin{table*}[t]
\begin{center}
\begin{tabular}{|c||c  |c |c |c  ||c |}

\hline

& El Ghazi 
& \multicolumn{2}{c|}{AlloyPE}
& \crs/\alloytosmt
&  \\
Aspect
& \cite{ElGhazi2010}
& \cite{ElGhazi2011a}, \cite{ElGhazi2014}
& \cite{ElGhazi2014}
& \cite{Meng2017}, \cite{Mohamed2019}
& \portus/\fortress \\
\hline
\hline 
Target
& Yices 
& SMT-LIB
& SMT
& CVC4 native
& SMT-LIB 
\\

&  
& (AUFLIA)
& (QBVF)
& (finite relations)
& 
\\
\hline
Sorts 
& multi-sorted
& multi-sorted
& bit vectors
& one sort
& multi-sorted 
\\
\hline
Scope
& mostly
& unbounded
& \textbf{bounded}
& unbounded
& \textbf{bounded}
\\

& unbounded
& 
& 
& 
& 
\\
\hline

Evaluation
& 9 
& 20~\cite{ElGhazi2011a}
& 4
& 40~\cite{Meng2017}
& 337 (correctness)
\\
(\# queries)
& 
& 6~\cite{ElGhazi2014}
& 
& 
& \numsupport\ (performance)

\\

\hline
\end{tabular}
\end{center}
\caption{Comparison of translations to SMT.} 
\label{tab:comparison}
\end{table*}

There have been several efforts to translate Alloy to SMT solvers. 
Our approach is distinct for its focus on using an SMT solver for 
finite model finding for the scopes provided in an Alloy model, 
thus keeping the problem completely decidable and 
equivalent to the original Alloy query.
An advantage of relying on EUF 
is that 
is widely supported in SMT solvers. Our 
approach will benefit from performance improvements in EUF solvers.
\portus\ is also novel in how it optimizes  
scalars and functions,
signature hierarchy together with
non-exact scopes, 
set cardinality, 
and how it uses
symmetry breaking (statically within an SMT solver) 
for the ordering module.  
\portus{} covers almost all Alloy constructs (including libraries) and is
fully integrated into the \alloyanalyzer.  Additionally,
we extensively evaluate our approach for completeness, correctness and performance.

\textbf{Translations from Alloy to SMT.} 
Table~\ref{tab:comparison} provides an overview of the related work on translating Alloy to an SMT solver. The number of queries is the number of queries evaluated (but not necessarily supported or completed by the tool) in the citation. First, we briefly present these works. Then, we discuss a comparison decomposed by Alloy language feature between our approach and prior work.

\textbf{Overview.} El Ghazi~\myetal~\cite{ElGhazi2010} report on a case study where three models (nine queries in total) of an address book in Alloy are translated by hand to the Yices SMT solver~\cite{Dutertre2014} for mostly unbounded verification.  
Since Yices allows subsorts, Alloy subsets are expressed using a combination of predicates and subsorts.  
Signatures used in transitive closure operations must have finite sorts.
Yices cannot complete evaluation over multiple transitive closures even with relatively small finite scopes in the examples.

Extending this work, El Ghazi~\myetal~\cite{ElGhazi2011a} present an automatic translation from Alloy to SMT-LIB (AUFLIA subset) for completely unbounded analysis where the problem is undecidable and thus
the SMT solver may not complete.  This translation is used in the tool \alloype~\cite{ElGhazi2014} for two of its three solving
strategies for Alloy: 1) SMT-based unbounded verification and 2) SMT-based bounded verification with a translation to quantified bit vectors (QBVF). The third method included in the tool \alloype\ is interactive-theorem-prover-based full verification using the \key{} theorem prover (based on~\cite{Ulbrich2012}), which is discussed later in this section. 
El Ghazi~\myetal~\cite{ElGhazi2011a} evaluate 20 queries where three produced unsound counterexamples due to an approximation of transitive closure and there is one timeout.
El Ghazi~\myetal~\cite{ElGhazi2014} (\alloype) evaluate the bounded method on 
six queries (all solved) and the unbounded method on four queries (with two timeouts).
Meng \myetal~\cite{Meng2017} report that \alloype\ fails on a number of Alloy models due to unsupported Alloy constructs or internal errors.

Astra~\cite{Abbassi2018a} is a precursor to \portus\ (which is why it is not included in Table~\ref{tab:comparison}) to 
translate from Alloy's \kodkod\ interface 
to \fortress{}. Astra 
translates only a very limited subset of the Alloy language 
and only supports analysis of
an exact scope size.
Astra uses a bottom-up approach to translation, which requires the creation of many auxiliary functions and extra quantifiers, causing bottlenecks in efficiency, which we avoid in \portus\ with a top-down approach.
Astra starts from the \kodkod{} model and thus requires
reverse-engineering steps to determine the set hierarchy, which we avoid by starting from the Alloy model.

Bansal~\myetal~\cite{Bansal2018} present a calculus and decision procedure (implemented in \cvcfour) for the decidable theory of flat finite (but unbounded) sets of individual elements (not relations) with cardinality and set membership.
This decision procedure is extended in Meng~\myetal~\cite{Meng2017} to handle formulas with relational operators such as join and transitive closure, and set comprehension. 
In the tool called \crs, they implement a fairly direct translation from Alloy's core language to this theory in CVC4's native language that uses
one sort for all Alloy atoms except for integers.
Meng~\myetal~\cite{Meng2017} report that on 40 queries, 20 are solved by their approach, but their approach is slower than the \alloyanalyzer\ for satisfiable problems. 
Mohamed~\myetal~\cite{Mohamed2019} report on an integration of this approach with the \alloyanalyzer, called \alloytosmt.

\qalloy~\cite{Silva2022} is an extension to Alloy (which is why it is not included in Table~\ref{tab:comparison}) that adds quantities to tuples of a relation implicitly rather than having integers in explicit positions in a tuple of a relation.  A tuple of atoms in Alloy has a quantity associated with it (rather than it just being present or absent in the relation). The advantages of a quantitative relation are that it supports multi-sets, typechecking can catch unit comparison errors, and there is a cleaner notation for mathematical calculations. 
Reasoning about quantitative relations is supported via a similar translation as found in \kodkod. However, membership of a tuple in a quantitative relation is not just a Boolean value (as in \kodkod) but now an integer.  Thus, an SMT solver is used to handle unbounded values for these quantitative relations.

\textbf{Feature Comparison.} As can be seen in Table~\ref{tab:comparison}, most of the related work on translations of Alloy to SMT solvers focus on unbounded verification with the exception of \alloype\ to QBVF~\cite{ElGhazi2014}. 
As a consequence, no other work has had to deal with translating scopes on subsignatures or non-exact scopes.
These works differ in whether they represent Alloy's signatures using multiple sorts or a single sort.  In \portus, we use multiple sorts in our partition sort policy because of the symmetry breaking schemes in \fortress, which  benefit performance.  Our method 
for merging sorts goes beyond what is needed in Edwards \myetal~\cite{Edwards2004a}.
\alloype\ to QBVF~\cite{ElGhazi2014} represent each sort using a fixed-size bit vector based on its scope.
Additionally, \portus\ supports bounded integers using SMT-LIB integers with an overflow-preventing bounding scheme in \fortress. 
In~\cite{Meng2017}, integers are represented as a finite set of a new sort called `uninterpreted int' with a mapping from that sort to unbounded integers.

Some existing methods recognize scalars in the Alloy model as scalars in SMT-LIB/FOL (\myeg~\cite{ElGhazi2011a,ElGhazi2014,Arkoudas2003}) but none of these related efforts have a systematic scheme for recognizing combinations of scalars in formulas as scalars, which is a contribution of~\portus.
Total and partial functions are represented using a variety of methods. In~\cite{ElGhazi2010}, an extra constant is introduced to be a `non value' for a partial function represented as a total function, which requires custom translation for uses of the partial function.  Other methods create a predicate and add axioms to limit the predicate to be a total function. In~\cite{ElGhazi2011a}, five patterns are introduced for simplifying formulas that include variations of quantification, functions, and equality.  We present a systematic method to recognize total and partial functions and their combinations as scalars in formulas to improve performance.  Our sort resolvants and membership predicates systematically handle the use of subsignatures in formulas.

Related efforts use a variety of methods to translate Alloy set operations ranging from axiomatizing the meaning of uninterpreted predicates for the Alloy kernel operations, such as union, intersection, and join (\myeg~\cite{ElGhazi2010,Ulbrich2012,Arkoudas2003}), to translating $(x_1,\dots,x_n) \in \alloytext{e}$ directly to MSFOL formulas as we use in \portus{} (\myeg~\cite{ElGhazi2011a,ElGhazi2014}).
Beyond the Alloy kernel, \portus\ translates domain/range restriction and set override, which are not present in related work.  Additionally, we translate set comprehension directly to MSFOL.  The only other work to support set comprehension is \crs~\cite{Meng2017}, which translates it directly to its operation in its theory of sets.

Previous work to support set cardinality uses a mapping from tuples in a relation to integers~\cite{ElGhazi2011a}.
In the approach of~\cite{Meng2017}, set cardinality in its theory of sets can only be compared to a constant number.  We present novel approaches in both our general translation for set cardinality and its optimization as  
constants-based scope axioms.
There are alternative approaches to handling the cardinality operators.  Cardinality can be expressed using sufficient first-order formulas introduced by Leino et al.~\cite{Leino2009} without finite bounds for comprehensions to prove some verification conditions.

\portus\ supports transitive closure in the same way as~\cite{ElGhazi2010}  using iterative squaring. Novelly, our translation supports trailing arguments for closures, which is needed to translate some models.
For closure over an unbounded sort, El Ghazi \myetal~\cite{ElGhazi2011a} use inductive definitions, which can cause spurious counterexamples. El Ghazi~\myetal~\cite{ElGhazi2015} extend this approach to transitive closure by automatically detecting
invariants iteratively to tighten the axiomatization of transitive closure over unbounded scopes. 

Additionally, the following features of Alloy that \portus\ supports are not mentioned in any other related work on translations from Alloy to SMT:
\begin{itemize}
\item signature merging that does not result in the same, single sort being used for all signatures, \myeg{} quantification over \alloy{A+B} (handled through splitting quantification when possible and sort merging when needed), with the exception of not allowing mixing \alloy{univ} and integers;
\item fields that refer to other fields in the same signature;
\item symmetry breaking for the ordering module to improve performance; 
\item if-then-else formulas;
\item \alloy{lone}/\alloy{one} formula expressions; and 
\item declaration formulas (\myeg\ \alloy{f in A one->one B}). 
\end{itemize}
While some of the above features translate into an Alloy kernel language, \portus\ provides translations for them directly to apply more optimizations.
Finally, \portus\ is the only SMT-based backend fully integrated within the \alloyanalyzer, and our evaluation is more extensive than any related efforts.

\textbf{Translations from Alloy to theorem provers.} 
There have been a few efforts to link Alloy with interactive theory proving tools.  
In these cases,
optimizing the translation for performance is less needed.
\prioni~\cite{Arkoudas2003} translates Alloy to the MSFOL of the Athena proof language for reasoning about sorts of unbound scope.    Alloy relations and relational operations are axiomatized and the key contribution is that through Athena, induction principles for structures can be created and used.
\kelloy~\cite{Ulbrich2012} is a translation of Alloy to the \key{} theorem prover~\cite{Beckert2007} based on many-sorted first-order logic.  In this translation, all Alloy fields are represented as relations and transitive closure and set cardinality are represented similarly to~\cite{ElGhazi2011a}.  Set operations are mapped to predicates of similar names, and the meaning of these predicates in terms of quantifiers become rewrite rules to be used by the theorem prover.
Users can interact with the theorem prover to help with quantifier instantiations, \myetc\
\kelloy\ uses finiteness predicates to state that a signature is finite.
Twelve of twenty-two assertions in ten models from the \alloyanalyzer\ distribution were proven automatically within 30 seconds.
The tool \dynamite~\cite{Moscato2014} provides a front-end to a shallow embedding of Alloy in \pvs~\cite{Owre1992} for proofs for unbounded scopes via fork algebras, which have similar operators to Alloy.  The proofs are interactive, but as a front-end, custom Alloy proof rules are provided that use the \alloyanalyzer\ at finite scope to help check introduced lemmas or search for witnesses to existential quantifiers.

\textbf{Translations from Alloy to other languages.} 
There are translations from Alloy to other formal languages.
Alloy 6 
includes support for the description of mutable behaviour and temporal properties in LTL and past LTL to constrain the possible behaviours of the model~\cite{Macedo2022}.   
Alloy 6 can translate an Alloy behavioural model to the SMV language for input to \nusmv~\cite{Cimatti2002} and \nuxmv~\cite{Cavada2014}.  In this translation, the finite relations of the Alloy model are represented in propositional logic using a Boolean variable for each tuple in a relation.
Krings \myetal~\cite{Krings2018} translate Alloy to B~\cite{Abrial1996}.  
As B has language constructs for sets, partial functions, and cardinality, the translation is fairly direct, but does not support libraries except the ordering module.  
They find that mapping the ordering module to integers is 
more efficient than using B sequences because B supports integers natively.
They test their translation on the river crossing and n-queens models, finding 
that sometimes verification with the B constraint solver \prob~\cite{Leuschel2008} is faster than Alloy.  
\alloytojml~\cite{Grunwald2014} translates Alloy specifications treating the heap as a relation to the Java Modelling Language~\cite{Leavens2008} for annotations of pre/post conditions of Java programs. Verification is done using the \key\ theorem prover~\cite{Beckert2007}.

\textbf{Finite Model Finders.} There are alternatives to using the \fortress{} library for finite model finding of translated Alloy models.
In Reynolds \myetal~\cite{Reynolds2013, Reynolds2013c, Reynolds2017}, a finite model finding decision procedure is added to \cvcfour\ that checks for satisfying instances of a model at gradually increasing scopes by looking at ground instances of the domain on-demand, rather than the use of domain constants and quantifier expansion (as \fortress{} does).  
Their results are promising, but at this time it is not a good match for expressing 
Alloy's scope and set cardinality constraints. 
When compared to \fortress{} in~\cite{Vakili2016b}, \cvcfour\ with finite model finding
did not perform well.

\allealle~\cite{Stoel2019a} is a relational model finding library, similar to \kodkod, which translates to an SMT formula in the logic of quantifier-free non-linear integer arithmetic, enabling it to support unbounded integers.
By using Codd's relation algebra as a base rather than Tarski's relational logic as \kodkod{} does, \allealle\ does not require data such as integers to be explicitly encoded in the domain.
\allealle\ also supports optimization criteria that change the order instances are returned.
\allealle\ translates to \zthree\ using only a single sort for non-integer atoms and does not apply any symmetry breaking optimizations.
In a comparison of six models, \allealle{} was compared to \kodkod\ without its symmetry breaking and \allealle{} was faster only on models containing integers and cardinality.

Reger \myetal{}~\cite{Reger2016} extend the \vampire~\cite{Kovacs2013} theorem prover to support finite model finding over MSFOL 
by encoding an MSFOL problem in propositional logic for a SAT solver in a way that takes advantage of multiple sorts, rather than using an SMT solver.
Their algorithm incrementally increases the sizes of the sorts to evaluate satisfiability over many model sizes.
They introduce techniques to determine relationships between the sort sizes in a problem so that fewer combinations of sort sizes need to be considered, including by analyzing returned proofs of unsatisfiability and by detecting properties of functions in the problem.
Subsorts are also detected and split to further reduce the generated formula size.
Their results show favourable performance compared to \cvcfour.

\section{Conclusion}

To the best of our knowledge, 
\portus\ provides the most complete method for linking Alloy to SMT solvers that is fully integrated into the \alloyanalyzer, including presenting instances from the tool within the Alloy evaluator and supporting the generation of a next instance of the 
model.\footnote{The \fortress\ solver supports the generation of a next instance.}
Our method translates Alloy models to the logic of equality with uninterpreted functions (EUF),
which has a decision procedure within most SMT solvers.
Our translation is novel in its separation of concerns between resolving sorts (which can handle signature-merging expressions in Alloy), recognizing scalars and combinations of scalars in join expressions, and translating almost all features of Alloy's formulas.  
Additionally, we use set membership predicates to handle Alloy's signature hierarchy and non-exact scopes and map set cardinality to integer operations, thus requiring no special decision procedure within the SMT solver.
Our technique is well-evaluated for completeness, correctness and performance, showing that \portus{} is competitive with \kodkod{}, and there is evidence that for a few language features, \portus{} may scale better than \kodkod{}. 
\portus{} forms a key component to creating a portfolio of solvers applicable to Alloy models.

Our work enables the use of SMT solvers for the analysis of Alloy models for a hybrid of bounded and unbounded signatures to obtain more general verification results.  We plan to add syntax to Alloy commands to allow the user to specify an unbound scope for certain signatures.  We also intend to investigate methods within the solver to recognize when 
a scope can be left unbound.

\ifCLASSOPTIONcompsoc
  \section*{Acknowledgments}
\else
  \section*{Acknowledgment}
\fi

We thank Michelle Zheng for her initial work on adding our new solver as an option in the \alloyanalyzer's graphical user interface.
We thank Ali Abbassi, Jose Serna,  Amin Bandali, Elias Eid, and Tamjid Hossain for their help in understanding Alloy. 
This research was supported in part by the Natural Sciences and Engineering Research Council of Canada (NSERC), Compute Ontario (https://www.computeontario.ca) and the Digital Research Alliance of Canada (https://www.alliancecan.ca).

\ifCLASSOPTIONcaptionsoff
  \newpage
\fi

\bibliographystyle{IEEEtran}

\bibliography{headings-short,from-nancy-shared,local}

\end{document}